\documentclass[prd,preprint,eqsecnum,nofootinbib,amsmath,amssymb,
               tightenlines,dvips]{revtex4}

\usepackage{graphicx}
\usepackage{bm}

\def\brem{bremsstrahlung }

\def\Tr{\: {\rm Tr} \:}
\def\M{{\cal M}}
\def\x{{\bf x}}
\def\p{{\bf p}}
\def\k{{\bf k}}
\def\q{{\bf q}}

\def\ca{C_{\rm A}}
\def\cf{C_{\rm F}}
\def\ch{C_{H}}

\def\nc{N_{\rm c}}
\def\nf{N_{\rm f}}
\def\mD{m_{\rm D}}

\def\PiT{\Pi_{\rm T}}
\def\PiL{\Pi_{\rm L}}
\def\alphas{\alpha_{\rm s}}
\def\st{\begin{equation}}
\def\stp{\end{equation}}
\def\bg{\begin{eqnarray}}
\def\nd{\end{eqnarray}}
\def\Eq#1{Eq.~(\ref{#1})}
\def\Sect#1{Section~\ref{#1}}
\def\Fig#1{Fig.~\ref{#1}}

\def\llangle{\left\langle}
\def\rrangle{\right\rangle}

\def\ito{{\scriptscriptstyle \rm Ito}}
\def\strat{{\scriptscriptstyle \rm Strat}}
\def\muD{\mu_{\scriptscriptstyle \rm D}}
\def\pperp{p^\perp}

\def\gsim{\mbox{~{\protect\raisebox{0.4ex}{$>$}}\hspace{-1.1em}
	{\protect\raisebox{-0.6ex}{$\sim$}}~}}
\def\lsim{\mbox{~{\protect\raisebox{0.4ex}{$<$}}\hspace{-1.1em}
	{\protect\raisebox{-0.6ex}{$\sim$}}~}}

\def\nott#1{\setbox0=\hbox{$#1$}                
   \dimen0=\wd0                                 
   \setbox1=\hbox{/} \dimen1=\wd1               
   \ifdim\dimen0>\dimen1                        
      \rlap{\hbox to \dimen0{\hfil/\hfil}}      
      #1                                        
   \else                                        
      \rlap{\hbox to \dimen1{\hfil$#1$\hfil}}   
      /                                         
   \fi}                                         %

\advance\parskip 1.9pt
\advance\voffset -0.2in

\newcommand{\ov}[1]{\overline{#1}}
\def\pperp{p^\perp}
\def\uperp{u^\perp}

\begin{document}

\vspace*{1cm}

\title{How Much do Heavy Quarks Thermalize in a Heavy Ion Collision?}

\author{Guy D. Moore}
\affiliation
    {%
    Department of Physics,
    McGill University, 3600 rue University,
    Montr\'{e}al QC H3A 2T8, Canada
    }%
\author{Derek Teaney}
\affiliation
    {%
    Department of Physics \& Astronomy,
    SUNY at Stony Brook,
    Stony Brook, NY 11764, USA
    }%

\date{\today}

\begin{abstract}

We investigate the thermalization of charm quarks in high energy heavy ion
collisions. To this end, we calculate the diffusion coefficient in the
perturbative Quark Gluon Plasma and relate it to collisional energy loss and
momentum broadening. We then use these transport properties to formulate a
Langevin model for the evolution of the heavy quark spectrum in the hot medium.
The model is strictly valid in the non-relativistic limit and for all
velocities $\gamma v < \alphas^{-1/2}$ to leading logarithm in $T/m_D$. The
corresponding Fokker-Planck equation  can be solved analytically for a Bjorken
expansion and the solution gives a simple estimate for the medium modifications
of the heavy quark spectrum as a function of the diffusion coefficient. Finally
we solve the Langevin equations numerically in a hydrodynamic simulation of the
heavy ion reaction. The results of this simulation are the medium modifications
of the charm spectrum $R_{AA}$ and the expected elliptic flow $v_2(p_T)$ as a
function of the diffusion coefficient.
\end{abstract}



\maketitle


\section {Introduction}

The experimental relativistic heavy ion program at RHIC
aims to measure the properties of the Quark Gluon Plasma (QGP) \cite{QM2004}.
One of the most exciting results from  RHIC  so far is
the large azimuthal anisotropy of light hadrons with respect
to the reaction plane, known as elliptic flow.   Elliptic flow has been
measured
as a function of  impact parameter, transverse momentum, rapidity and
particle type and is quantified with $v_2(p_T)$
\cite{Adams:2004bi,Adler:2002pu,Adcox:2002ms,Vale:2004dw,Back:2002gz},
\begin{eqnarray}
      v_2(p_T) &=&
      \frac{ \int\,d\phi\, \frac{dN}{p_T dp_T d\phi}\, \cos(2\phi)  }
           { \int\,d\phi\,  \frac{dN}{p_T dp_T d\phi}  } \; .
\end{eqnarray}
The observed elliptic flow is significantly larger than was originally expected
from kinetic calculations of quarks and gluons \cite{Molnar:2001ux},
but in fairly good agreement with simulations
based upon ideal hydrodynamics
\cite{Hirano:2004er,Teaney:2001av,Kolb:2000fh,Huovinen:2001cy}. This
result suggests that
the medium responds as a thermalized fluid and
that the transport mean free path is small
\cite{Molnar:2001ux,Molnar:2004yh,Teaney:2003pb}.

However, this interpretation of the elliptic flow results is not
universally accepted \cite{Arnold:2004ti,Molnar:2003ff}.
Hadronization may amplify the underlying partonic elliptic flow \cite{Lin:2001zk}. Indeed, parton coalescence is one mechanism which
may amplify the hadronic elliptic flow relative to its partonic constituents \cite{Molnar:2003ff, Fries:2003vb, Greco:2003xt, Greco:2003vf, Lin:2003jy}.
On physical grounds, it seems unlikely that
the typical mean free path is much smaller than a thermal wavelength,
$1/(2\pi T)$\,.
Indeed it has recently been conjectured that the hydrodynamic diffusion
coefficient $\eta/(e + p)$
is strictly larger than half a thermal wavelength \cite{Kovtun:2004de},
\[
           \frac{\eta}{e + p} > \frac{1}{4\pi T} \; .
\]
Here $\eta$ is shear viscosity, $(e +p)$ is the enthalpy, and
the ratio is a fundamental length scale in the QGP.
If  $\eta/(e+p)$ is significantly larger than this conjectured bound
hydrodynamics would not be a viable explanation for the
observed flow at RHIC.
In this work we will accept the hydrodynamic paradigm of the
RHIC results and study the correlated consequences of this interpretation.

Heavy quarks are a good probe of the transport properties of the medium.
Given an estimate of the light quark relaxation time $\sim \eta/(e+p)$,
the heavy quark relaxation time is
\[
            \tau_{R} \sim \frac{M}{T} \frac{\eta}{e+p} \; ,
\]
where $M$ is the mass of the heavy quark and $T$ is the temperature.
Thus,  with $M\approx 1.4\,\mbox{GeV}$ and
$T \approx 250\,\mbox{MeV}$, we expect that the charm equilibration
time is approximately $6$ times larger than the light quark equilibration
time. Since this factor is relatively large, we further expect that the elliptic
flow of charm quarks will be smaller than the flow of light hadrons.

In addition, the heavy quarks are produced with a power-law transverse
momentum  spectrum which deviates strongly from the thermal spectrum.
The relaxation time will control
the extent to which the initial power-law spectrum approaches the thermal spectrum.
Similarly, the relaxation time will control the extent to which
the charm quark will follow the underlying flow of the medium.
If the charm quark completely follows the flow of the medium then thermal
spectrum is actually quite close to the perturbative spectrum and it
may be difficult to distinguish these two cases \cite{Batsouli:2002qf,Greco:2003vf}.
Medium modifications of the heavy quark spectrum $R_{AA}$ flow will be studied
experimentally this year and will provide an experimental estimate of this
relaxation time \cite{Adams:2004fc,Averbeck:2004dv,Adler:2004ta,Kelly:2004qw,Adcox:2002cg}.

Most recent studies of the medium modifications of the charm spectrum
have computed the energy loss of a heavy quark by
gluon \brem \cite{Djordjevic:2003qk,Djordjevic:2003zk,Dokshitzer:2001zm,Armesto:2003jh}.  In weak coupling (which is the framework in
which all calculations have been performed),
\brem is the dominant energy loss mechanism only if the heavy quark is
ultra-relativistic, $\gamma v  \gg 1/g$.
(Similarly, for an electron traversing a hydrogen target,
\brem losses first exceed ionization losses
when
$\gamma v \simeq 700$ \cite{pdg_book}.)  For much of the
measured momentum range, the heavy quark is not ultra-relativistic,
$\gamma v \lsim 4$, and
in this case it is far from clear that
radiative energy loss dominates over collisional energy loss.
%
%

About two thirds of all heavy quarks  are produced
with $p \lsim M$, and therefore radiative energy loss
should be neglected when studying  bulk thermalization.
When $\gamma v \gsim 4$, calculations do suggest that radiation
dominates the ${\it average}$ energy loss rate
\cite{Mustafa:1997pm,GolamMustafa:1997id}. However, 
as has been repeatedly emphasized \cite{Baier:2001yt,Jeon:2003gi}, the
${\it average}$ energy loss 
is insufficient to describe the medium modifications of the spectrum
$R_{AA}$. Collisions have a different fluctuation spectrum than
radiation and therefore might contribute more  to the
suppression factor than was at first anticipated \cite{Mustafa:2003vh}.
Since we are primarily interested here in heavy quarks with typical
momenta $\gamma v \sim 1$, we will concentrate exclusively on elastic
collisions. 

Considering these points, we will re-examine collisional energy loss of a heavy
quark in the perturbative QGP. Our tools are Hard Thermal Loop (HTL)
perturbation
theory and a heavy quark expansion ($M\gg T$).  The average energy loss rate
was first computed by Braaten and Thoma \cite{Braaten:1991we}  and we will
independently verify their results.  (Recently this calculation was extended to 
anisotropic plasmas by Romatschke and Strickland \cite{Romatschke:2004au}.) We will also compute the rates of
longitudinal and transverse momentum broadening which are essential to a
complete calculation of the modification factor $R_{AA}$.
We will relate all of these rates to the diffusion coefficient which we
will compute.  (In principle the diffusion coefficient of a heavy quark could
have been gleaned from the results of Braaten and Thoma \cite{Braaten:1991we}
and Svetitsky \cite{Svetitsky:1987gq}.)  In perturbation theory, we can compare
the diffusion coefficient of the heavy quark to the hydrodynamic time scale
$\eta/(e +p)$ which was calculated previously \cite{Baymetal,AMY1,AMY6}.
Many of the
ambiguities of perturbation theory cancel in the ratio of transport
coefficients and we therefore hope to be able to extrapolate smoothly into the
non-perturbative domain.  Following this ideology, we express all of our
phenomenological results in terms of the diffusion coefficient 
which may ultimately be determined from lattice QCD calculations.

With these transport properties in hand, we adopt a Langevin model for the
equilibration of heavy quarks in heavy ion collisions.
The Langevin equations correctly describe the kinetics of a
heavy particle in a thermal medium and therefore naturally interpolate
between a hydrodynamic regime at small momentum and a
kinetic regime at large momentum.
The model is similar to
old work by Svetitsky \cite{Svetitsky:1987gq} and was later used without
fluctuations to estimate $R_{AA}$ \cite{GolamMustafa:1997id}.  The model
is strictly valid for non-relativistic quarks and for all velocities to leading logarithm in $T/m_D$.
The corresponding Fokker-Planck equation is solved analytically
in \Sect{Bjorken} for a Bjorken expansion.  The solution
provides a simple estimate for the modification factor $R_{AA}$  and further
elucidates the dynamics of equilibration.  Then we solve the Langevin equations
numerically in a hydrodynamic simulation of the heavy ion reaction. The results
of the simulation are the medium modification factor $R_{AA}$ and the
corresponding elliptic flow $v_2(p_T)$ as a function of the diffusion
coefficient.

Very recently, two papers have adopted a similar Langevin approach to address
charm equilibration \cite{vanHees:2004gq,Mustafa:2004dr}. The first of these
papers \cite{vanHees:2004gq} estimated possible non-perturbative contributions
to the drag coefficient which may arise from quasi-hadronic bound states in
the QGP. The second of these papers \cite{Mustafa:2004dr} estimated radiative
and collisional energy loss, and found that collisional loss is significant
even for rather energetic charm quarks, $E \sim 5-10\,\mbox{GeV}$. 

In addition, classical Boltzmann simulations by Molnar \cite{Molnar:2004ph} and
subsequent simulations by Zhang {\it et al.\ }\cite{Zhang:2005ni}  have studied how $v_2(p_T)$ and
studied how $v_2(p_T)$ and
$R_{AA}$ depend on the charm mean free path. As is 
discussed in \Sect{comparison}, the results of
Molnar's simulation are comparable to the Boltzmann-Langevin approach
adopted here.

Throughout, we will denote 4-vectors with capital letters $P,Q$
and use $\p,\q$ for their 3-vector
components, $p^0,q^0$ for their energy components, and $p,q$
for $|\p|,|\q|$.  Our metric convention is [--,+,+,+].

\section{Non-Relativistic Heavy Quarks in a Thermal Medium}
\label{diffusion:sect}

First consider thermal heavy quarks, $M \gg T$, with typical thermal
momentum $p \sim \sqrt{MT}$ and velocity $v \sim \sqrt{T/M} \ll 1$.
Since $p \gg T$ it takes many collisions to change the momentum
substantially. Even for hard collisions with momentum transfer $q\sim T$, it takes $\sim M/T$
collisions to change the momentum by a factor of order one.
Therefore it is a good approximation to
model the interaction of the heavy quark with the medium as uncorrelated momentum kicks.
The momentum of the heavy quark will evolve according to the macroscopic
Langevin equations \cite{Reif}
\begin{equation}
\label{newton}
\frac{dp_i}{dt} = \xi_i(t) - \eta_D p_i \; , \qquad
\langle \xi_i(t) \xi_j(t') \rangle = \kappa \delta_{ij}
	\delta(t-t') \; .
\end{equation}
Here $\eta_D$ is a momentum drag coefficient and $\xi_i(t)$ delivers random momentum kicks which are uncorrelated in time. $3\kappa$ is the mean
squared momentum transfer per unit time.  The solution of this
stochastic differential
equation  is
\begin{equation}
p_i(t) = \int^t_{-\infty} dt' e^{\eta_D(t'-t)} \xi_i(t') \; ,
\end{equation}
where we have assumed that $t \gg \eta_D^{-1}$.  The mean squared value of $p$ is
\begin{equation}
3MT = \langle p^2 \rangle = \int^0 dt_1 dt_2 e^{\eta_D(t_1+t_2)}
	\langle \xi_i(t_1) \xi_i(t_2) \rangle
= \frac{3 \kappa}{2\eta_D} \; ,
\end{equation}
and  therefore
\begin{equation}
\label{etad}
\eta_D = \frac{\kappa}{2 MT} \; .
\end{equation}

Now the diffusion constant in space, $D$, can be found by starting a
particle at $x=0$ at $t=0$ and finding the mean squared position at a
later time,
\begin{equation}
\langle x_i(t) x_j(t) \rangle = 2D t \delta_{ij} \quad \rightarrow \quad
6D t = \langle x^2(t) \rangle \; .
\end{equation}
Using the relation between position and momentum
\begin{equation}
x_i(t) = \int_0^t dt' \frac{p_i(t')}{M} \; ,
\end{equation}
we have
\begin{equation}
6D t = \int_0^t dt_1 \int_0^t dt_2 \frac{1}{M^2} \langle p(t_1) p(t_2)
	\rangle
     = \frac{6Tt}{M\eta_D} \; ,
\end{equation}
and therefore the diffusion constant is \cite{Reif}
\begin{equation}
D = \frac{T}{M \eta_D} = \frac{2 T^2}{\kappa} \; .
\label{eq:D}
\end{equation}
In the next paragraphs we will determine the mean squared momentum transfer per unit time, $3\kappa$, and then the diffusion coefficient.

The only $2 \leftrightarrow 2$ scattering processes
are $qH \to qH$ ($H$ the heavy quark) and $gH \to gH$.  The  $qH \to qH$
process only occurs by $t$ channel gluon exchange.
Since in the rest frame of the plasma the
Compton amplitude is  suppressed by $v^2 \sim
T/M$ (see Appendix \ref{appendixA}),
the $gH \to gH$ process is also dominated by $t$ channel gluon exchange.
Kinematics demand that the transfer momentum be spatial up to $O(v)$
corrections, so the Hard Thermal Loop correction on the transferred gluon
is described simply by Debye screening.  Writing the incoming and
outgoing momenta of the thermal particle as $\k$ and $\k'$, and taking
that particle's dispersion relation to be ultra-relativistic, the
scattering matrix elements squared, summed on the colors and spins of
the incoming thermal particles and over all quantum numbers of the final
state particles, are (see Appendix \ref{appendixA}),
\begin{eqnarray}
|\M|^2_{\rm quark} & = & \left[2 \frac{\ch g^4}{2} \right] 
	\, 16 M^2 k_0^2 \,( 1 + \cos \theta_{\k\k'} )
	\frac{1}{(q^2 + \mD^2)^2}
	\; , \nonumber \\
|\M|^2_{\rm gluon} & = & \left[\nc \ch g^4 \right] 
	\, 16 M^2 k_0^2 \, ( 1 + \cos^2 \theta_{\k\k'} )
	\frac{1}{(q^2 + \mD^2)^2}
	\; .
\label{eq:Msq1}
\end{eqnarray}
$\ch=\cf$ denotes the color Casimir of the heavy quark and an 
extra factor of 2 for the quark case accounts for anti-quarks.

The mean squared momentum transfer per unit time is $3\kappa$.
To compute this quantity the matrix elements must
be integrated over the incoming momentum $\k$ and outgoing momenta
$\k'$, $\p'$, weighted by the appropriate statistical functions and by
the squared momentum transfer $q^2 \equiv (\k-\k')^2$.
The mean squared momentum transfer per unit time ($3\kappa$) is then
\bg
\label{eq:kappa1}
3 \kappa & = & \frac{1}{2M} \int \frac{d^3 \k d^3 \k' d^3 \p'}{(2\pi)^9
	8 k^0 k'{}^0 M} (2\pi)^3 \delta^3(\p + \k' - \p'-\k) 2\pi
	\delta(k'-k) \q^2 \times \nonumber \\
	&& \qquad \qquad \times
	\left[ N_f |\M|^2_{\rm quark} n_f(k) (1{-}n_f(k'))
	+|\M|^2_{\rm gluon} n_b(k) (1{+}n_b(k')) \right] \; .
\nd
In writing this formula
we have used the non-relativistic limit $p^0 = p'{}^0 = M$.
It is convenient to shift the $\p'$
integration to an integral over the momentum transfer $\q \equiv \p' - \p$
; the momentum conserving delta function becomes $\delta^3(\k' - \q -\k)$.
The appendix shows how a simple change of variables and an expansion in
$\mD^2 \ll T^2$ makes these integrals relatively straightforward.
Using the relation between $\kappa$ and the diffusion coefficient \Eq{eq:D},
we find
\st
D = \frac{36 \pi}{\ch g^4 T}
	\left[ \nc \left( \ln\frac{2T}{\mD}+\frac{1}{2} - \gamma_{\rm
	E} + \frac{\zeta'(2)}{\zeta(2)} \right)
	+ \frac{\nf}{2}\left( \ln\frac{4T}{\mD}+\frac{1}{2} -
	\gamma_{\rm E} + \frac{\zeta'(2)}{\zeta(2)} \right) \right]^{-1}
	.
\label{eq:D_heavy}
\stp
This formula for the diffusion coefficient could have been extracted
by combining the calculations of Braaten and Thoma \cite{Braaten:1991we}
and Svetitsky \cite{Svetitsky:1987gq}.
Corrections to this expression are suppressed by at least one power of
$\sqrt\alphas$;  we expect corrections at that level.

\begin{figure}
\begin{center}
\includegraphics[height=3.0in,width=3.0in]{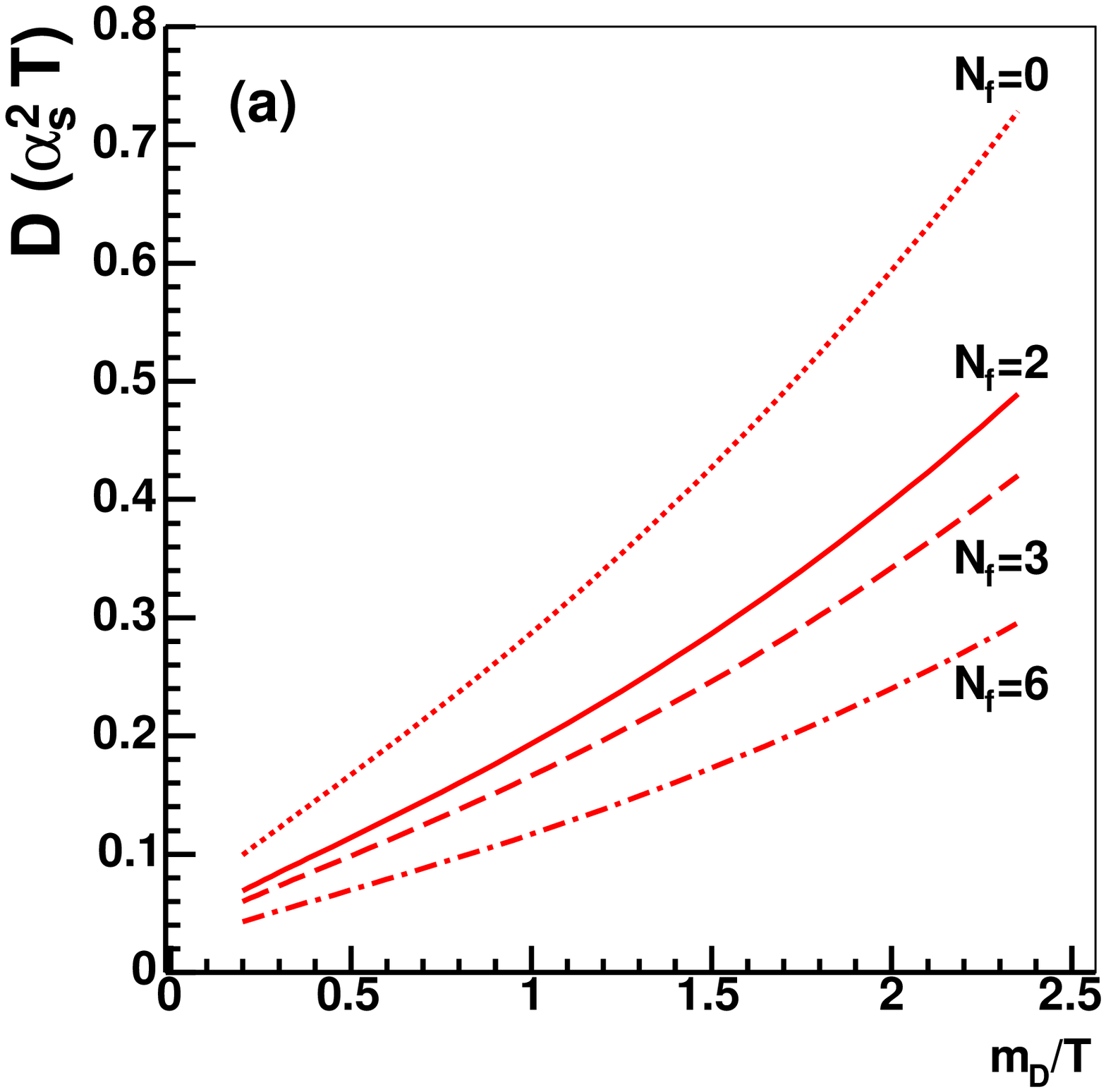}
\includegraphics[height=3.0in,width=3.0in]{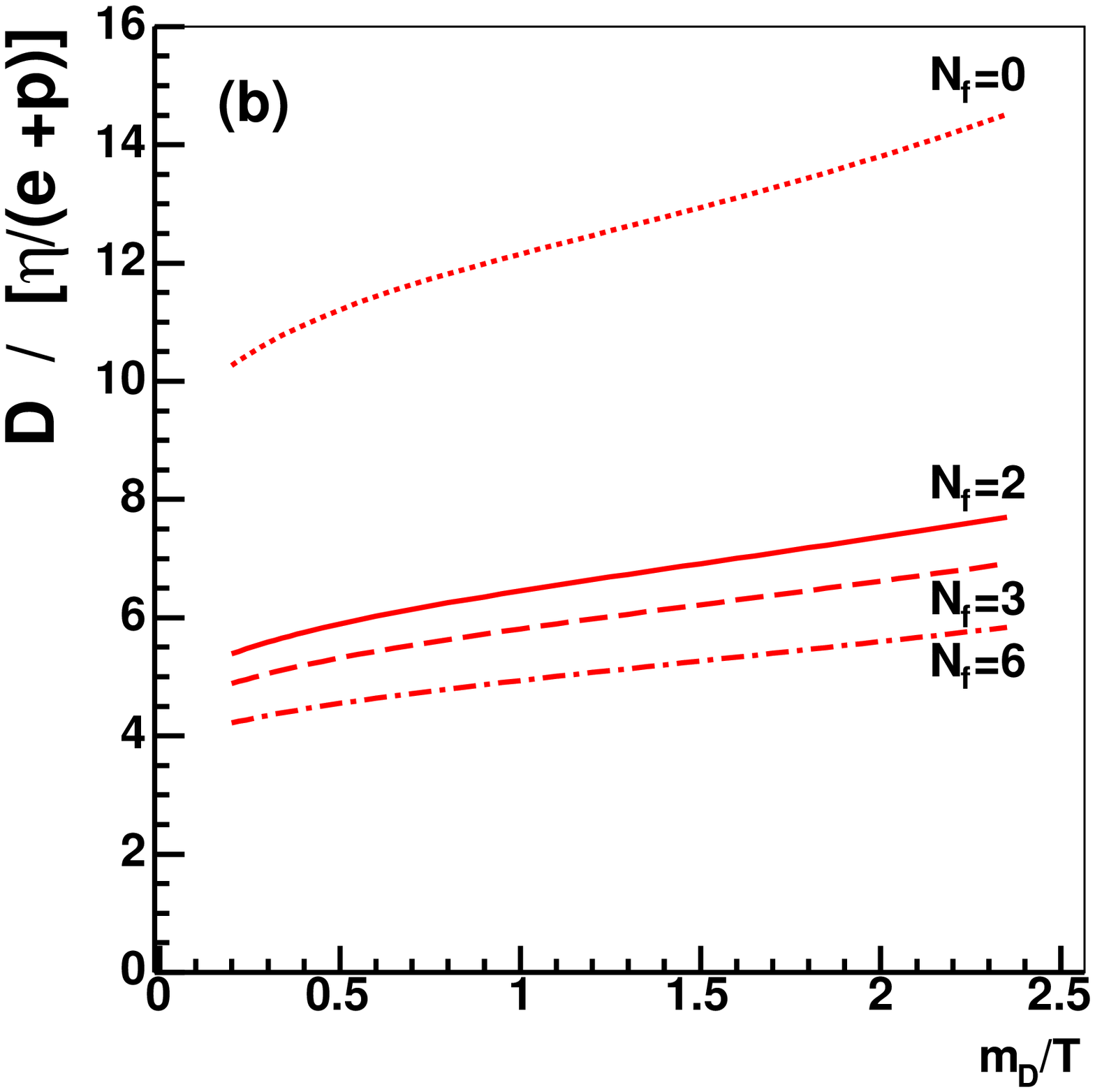}
\caption{
(Color online) (a) The diffusion coefficient of a heavy quark in a QGP with
$N_f$ flavors of light quarks. (b) The ratio of the diffusion coefficient
of a heavy quark to the hydrodynamic diffusion coefficient $\eta/(e+p)$.
}
\label{diffusion}
\end{center}
\end{figure}

This expression for the diffusion coefficient  is based
on a small $\mD$ expansion and therefore becomes unreliable
when the Debye mass becomes large.
Rather than using a small $\mD$ expansion we can numerically
integrate \Eq{eq:kappa1} to determine the
the diffusion coefficient.  Something of this
sort is necessary to deal with large values of $\mD$,
but we emphasize that, at
large values of $\mD$, the procedure is ad-hoc and should be considered only a
qualitative guide. The $\mD$ expansion would be a good approximation
if the coupling were truly small.
The numerical evaluation of the diffusion coefficient is
illustrated in Fig.~\ref{diffusion}(a)
as a function of $\mD$.

As discussed in the introduction, the time scale for
equilibration $\tau \sim \frac{M}{T} \frac{\eta}{e + p} $.
Now let us make this more concrete.
The rate of equilibration is given by $\eta_{D}^{-1}$,
as can be intuited from the Langevin equations, \Eq{newton}, or
from the analysis of \Sect{Bjorken}. $\eta_{D}^{-1}$ is
directly related to the diffusion coefficient via
\begin{eqnarray}
\frac{1} {\eta_D}  = \frac{M}{T}D   \; .
\end{eqnarray}
Taking $\alphas \approx 0.5$,  $\mD/T\approx 1.5$, and $M/T \approx 7$
we estimate that the diffusion coefficient is $D\approx \frac{1.0}{T} \approx
\frac{6}{2\pi T} $.
The relaxation time is then $\eta_D^{-1} \approx 6.7/T$.

It is useful to compare this timescale with other  hydrodynamic
relaxation times in the QGP. The damping rate of sound waves in
the QGP is controlled by $\eta/(e+p)$ where $\eta$ is
the shear viscosity  and
$e+p$ is the enthalpy.
Dividing by this ratio we have
\begin{eqnarray}
\frac{1} {\eta_D}  = \frac{M}{T} \; \frac{\eta}{e + p} \;
\times  \left[ \frac{D} {  \eta/(e + p)  } \right] \; .
\end{eqnarray}
The quantity in square brackets is an estimate of
the ratio between the heavy quark and hydrodynamic relaxation times
and is illustrated in Fig.~\ref{diffusion}(b) as a
function of $\mD/T$. Here the shear viscosity at leading
order has been taken from \cite{AMY6}.
As seen from the figure, for $N_f\approx3$
the diffusion coefficient is $\approx 6$
times larger than hydrodynamic scale $\eta/(e + p)$.

Both the calculation of the diffusion coefficient $D$ 
and of the shear viscosity $\eta$ are
plagued by ambiguities when the coupling is strong, $\alphas
> 0.2$.  In particular, in both processes it is unclear
how to screen the plasma when $\mD$ is large and how to estimate
the form and size of subleading $\alphas$ corrections.  These
issues have been discussed in 
\cite{AMY6}, but  were not resolved.  
However, since both processes are dominated by $t$-channel
gluon exchange, the ambiguities in the calculation
should  largely cancel in the ratio of transport coefficients.
We therefore hope that the computed factor of $\approx 6$  for $D/[\eta/(e+p)]$
is largely independent of its perturbative assumptions.
Thus, with $\frac{M}{T} \approx 7$, $T\approx 200\,\mbox{MeV}$ and
an optimistic estimate of the shear viscosity,
$\frac{\eta}{e + p} =  1/(6T)$, we estimate that the
charm quark thermalization time is of order $\approx 7\,\mbox{fm}$.
This time scale should be compared with the
time scale for the development of elliptic flow  $\approx 4\,\mbox{fm}$.

\section{Energy loss}

Since the initial distribution of charm quarks is much ``harder'' than
a thermalized spectrum, it is important to study how higher energy heavy
quarks, with $\gamma v \sim 1$, lose their energy in the thermal
medium.  We now turn to a discussion of this problem.  Our discussion
differs from most recent literature in that we take the dominant energy
loss mechanism to be elastic scattering.

\subsection{Why $2\leftrightarrow 2$ dominates for $\gamma v \sim 1$}

When the coupling is small, \brem dominates
the energy loss rate for very fast particles (with $\gamma v \sim 1/g$)
while collisions  dominate the rate for moderately relativistic particles with
$\gamma v \sim 1$.

To see why,
consider a heavy particle of momentum $v$, undergoing a scattering.
Call its momentum $P$, with $p\equiv |\p| = v p^0$.  For
energy loss in a condensed medium, the scattering is typically off a
nucleus, so the transferred 4-momentum $Q^\mu$ is purely spatial.  In a
thermal medium like the QGP, the scattering is off a relativistic
particle at a random angle with respect to the heavy particle; the
kinematics of the other particle requires that $Q^\mu$ be spacelike,
but $q^0$ can be $\sim q\equiv |\q|$.  Kinematics requires that the
lightlike, bremmed particle's momentum $K$ satisfy $(K+P')=(Q+P)$, with
$P'$ the final particle momentum.  $K$ can be the largest if it is
perfectly collinear with $P$ and if $Q$ is anti-collinear; in this case,
energy conservation reads $k^0 = p^0 - p'{}^0 + q^0 = v (p-p') + q^0$,
while momentum conservation reads $k=k^0 = p-p'-q$.  For the spatial
(Coulomb) scattering case, $k \leq vq/(1{-}v)$, while for the QGP case,
$k \leq (vq {+} q^0)/(1{-}v)$, so the actual energy loss is
$p^0-p'{}^0\leq v(q{+}q^0)/(1{-}v)$.  In either case, for $\gamma v \sim
1$, the energy of a bremmed particle is at best of order of the transfer
momentum in the scattering.

The energy loss in a scattering process without \brem is
$q^0$.  When $q^0 \sim T$ the plasma temperature, phase space favors
events with $q^0$ negative and of order $q$.  Since the rate of
\brem is suppressed by an additional factor of $\alphas$,
and the typical energy loss is the same, the contribution of
\brem to energy loss is smaller by a factor of $\alphas$.  In
the soft region, $q \sim g T$, there is an approximate cancellation
between energy losing and energy gaining scatterings, and the dominance
of $2\leftrightarrow 2$ processes is only by $\sqrt\alphas$, but they
remain dominant.  Softer momentum transfers are screened by the plasma
and do not play an important role in energy loss.

Very similar arguments apply for electromagnetic energy loss of a high
energy particle in a medium, except that the scale $g T$ is played by
the atomic radius and the $2\leftrightarrow 2$ processes do not have any
cancellation between energy losing and gaining processes; for $\gamma v
\sim 1$, \brem energy losses are subdominant to ionization
losses (due to elastic scattering with electrons, transferring more
energy than the electron's binding energy) by $Z \alpha$.

As the particle's energy increases, the importance of \brem
also increases.  While the \brem cross-section remains suppressed by a power
of $\alphas$, the amount of energy which can be lost in a single event
increases.  As we will see, the rate of energy loss by $2
\leftrightarrow 2$ processes only increases logarithmically in $\gamma
v$.  Bremsstrahlung energy losses typically rise almost linearly in
$\gamma v$, and so become dominant at $\gamma v \sim 1/Z \alpha$ for
electromagnetic processes and at $\gamma v \sim 1/\sqrt{\alphas}$ for
heavy quark energy loss in the QGP.  Since only a small tail of heavy
quarks have $\gamma v \gg 1$, we will only treat $2 \leftrightarrow 2$
processes in this work.

\subsection{Relativistic heavy quarks; momentum loss and momentum
diffusion}
\label{eloss-sect}

First consider a heavy quark in a static medium with $p \gg T$
and velocity $\gamma v \sim 1$.
It takes $\sim p/T$ collisions to change the momentum of
the heavy quark by a factor of order one.  The time between
hard collisions is $\sim 1/(g^4 T)$ and thus
 the equilibration time scale  a heavy quark is of
order $\sim (p/T)\, 1/(g^4 T)$.

Next consider how a heavy quark with momentum $p \gg T$  changes over
a time interval $\Delta t$
which is long
compared to the timescale of medium correlations but short compared to
the time scale of heavy quark thermalization,
\st
     \frac{1}{g^4 T} \ll \Delta t \ll \frac{p}{T} \frac{1}{g^4 T}   \;.
\stp
For $\Delta t$ large compared to  $1/(g^4 T)$ the number of collisions $N$
is large $\sim (\Delta t) (g^4 T)$. On the other hand the momentum of the heavy particle has scarcely changed since the total momentum
transfered
$\Delta p$ is of order  $\Delta p \sim  T (\Delta t) (g^4 T) $ which is
small compared to $p$. This
means that the probability distribution for a given momentum transfer from
a single collision is approximately constant over this time period.
The accumulated momentum transfer is a sum of $N$ such collisions. The sum of
a large number of momentum transfers drawn from an identical
probability distribution is approximately Gaussian plus corrections
which go as $1/N$.
In the next $\Delta t$ time interval the process repeats itself independently.
Thus we can write down
a macroscopic equation of motion for the
the heavy quark moving in the $z$ direction,
\begin{eqnarray}
    \frac{d }{dt} \llangle p \rrangle  &\equiv& - \eta_D(p) p  
	\; , \nonumber \\
    \frac{1}{2} \frac{d }{dt} \llangle (\Delta p_T)^2 \rrangle 
	&\equiv& \kappa_T(p) \; ,  \nonumber \\
    \frac{d }{dt} \llangle (\Delta p_z)^2 \rrangle &\equiv& \kappa_L(p) 
	\nonumber \; .
\end{eqnarray}
Here $\llangle (\Delta p_T)^2 \rrangle = \llangle p_T^2 \rrangle$
is the variance of the momentum distribution
transverse to the direction of the heavy quark, $\llangle (\Delta p_z)^2 \rrangle =(p_{z} - \llangle p_z\rrangle)^2$ is the variance of
the momentum distribution in the direction
parallel to the direction of the quark, and the time derivatives
are understood to act only on a time scale of order $\Delta t$.
The factor of $\frac{1}{2}$ in the transverse fluctuations has been
inserted  because there are two perpendicular directions.
The functions $\eta_D$, $\kappa_T$, and $\kappa_L$ encode the
average momentum loss and the transverse and longitudinal fluctuations.

Now we will compute  these coefficients using  kinetic theory.
First we will compute the mean rate of momentum loss.
This is most easily done by computing the
energy loss rate, $dp^0/dt$, which is related to the momentum loss by $dp^0/dt = v dp/dt$.
The energy loss rate is found by multiplying the scattering rate with transfer
energy $q^0$, schematically,
\st
\frac{dp}{dt} = \frac{1}{v} \int_{k,q} |\M|^2 q^0
	f[k] (1\pm f[k{-}q^0]) \; .
\stp
The factor $q^0$ is not enough to render this IR convergent, but there
are cancellations between $q^0>0$ and $q^0<0$ contributions.  It is
easiest to account for these by averaging in the integrand over the
process with $k$ incoming and $k'$ outgoing, and the process with $k'$
incoming, $k$ outgoing, and opposite $Q^\mu$.  Because we take $p \gg k$
the kinematics and
matrix element are the same, but the population
functions differ, yielding
\st
\frac{dp}{dt} = \frac{1}{2v} \int_{k,q} |\M|^2 q^0
\left\{ f[k] (1\pm f[k{-}q^0]) - f[k{-}q^0] (1\pm f[k]) \right\} \; .
\label{eq:fs}
\stp
The population function here is $( e^{k/T} - e^{(k-q^0)/T} ) f[k] f[k-q^0]$,
which vanishes at small $q^0$ and makes the integral well behaved.

Similarly, to compute the rate of transverse momentum broadening, we weight
the  transition rate $|\M|^2$ with the square of the transverse
momentum transfer
\st
\frac{d}{dt}  \llangle (\Delta p_T)^2 \rrangle  =
\int_{k,q} |\M|^2 q_{T}^2 f[k] (1\pm f[k{-}q^0]) \; .
\stp
Here the integral will be convergent
and we do not need to symmetrize over
forward and backward collisions.  Finally the
rate of longitudinal momentum  broadening
is
\st
\frac{d}{dt}  \llangle (\Delta p_z)^2 \rrangle  =
  \int_{k,q} |\M|^2 q_{z}^2 f[k] (1\pm f[k{-}q^0]) \; .
\stp
Again this integral is convergent.  Thus to compute
the transport coefficients $\eta_D$, $\kappa_T$, $\kappa_L$
we need to specify the matrix elements and perform (numerically) the
phase space integrals.

The covariant expressions for the scattering matrix elements in
\Eq{eq:Msq1} (still summed over spins and colors of the bath particle),
in vacuum, are
\bg
\label{eq:Msq2}
|\M|^2_{\rm quark} & = & \left[ 2\frac{\ch g^4}{2}
	\right] 16 \left[ 2 \frac{(P\cdot K)^2}{Q^4} -
	\frac{M^2}{2Q^2} \right] \; ,
	\nonumber \\
|\M |^2_{\rm gluon} & = & \left[ \nc \ch g^4 \right] 16
	\left[ 2 \frac{(P\cdot K)^2}{Q^4} -
	\frac{M^2}{Q^2} + \frac{M^4}{4(P\cdot K)^2} \right] \; ,
\nd
where $\ch=\cf$ is the quadratic Casimir of the heavy quark, $P$ is the heavy particle 4-momentum, $K$ is the light
particle 4-momentum, $Q$ is the 4-momentum transfer, and we use [--,+,+,+] metric
convention.  The inclusion of Hard Thermal Loops is only necessary at
small $Q^2$, where the leading $1/Q^4$ term dominates; the details
appear in the appendix. This introduces the Debye mass into the
problem and thus the results will generally depend on the ratio
of $m_D/T$.

Analytic expressions for the
transport coefficients can be derived to leading logarithm in $T/m_D$.
Without the Hard Thermal Loop correction the phase space integrals
are logarithmically divergent at small $q^2$. The leading log
transport coefficient is found by evaluating the contribution
to the phase-space integrals  from the logarithmically
divergent region $\mD^2 \ll q^2 \ll T^2$.
This integral will give a log of $T/\mD$ times some function of $v$.
Further details  are given in the appendix. We find that, to
leading logarithm, the  transport coefficients are
\begin{eqnarray}
\label{dpdtll}
\frac{dp}{dt}
& \simeq &  v \left( N_c + \frac{N_f}{2} \right)
	\frac{\ch g^4 T^2}{24\pi} 
	\left( \frac{1}{v^2} - \frac{1{-}v^2}{2v^3}
	\ln\frac{1{+}v}{1{-}v} \right) \ln(T/\mD)
\; , \\
\kappa_T  &\simeq &    \left( N_c + \frac{N_f}{2} \right)
	\frac{\ch g^4 T^3}{12\pi} 
	\left( \frac{3}{2} - \frac{1}{2v^2} + \frac{(1{-}v^2)^2}{4v^3}
	\ln\frac{1{+}v}{1{-}v} \right) \ln(T/\mD)\; , \\
\kappa_L  &\simeq &  \frac{2 T}{v} \frac{dp}{dt} \; .
\label{eq:leadinglog}
\end{eqnarray}
We note that the relation $\kappa_L = \frac{2 T}{v} \frac{dp}{dt}$
is what is expected of the Boltzmann-Langevin approach.  Indeed,
the conclusion of the next section is that the Boltzmann-Langevin
approach is strictly valid only in a leading-log approximation
or in the non-relativistic limit.
For completeness, the constant under the logarithm in Eqs.~(\ref{dpdtll})--(\ref{eq:leadinglog}) is calculated in the appendix.

\begin{figure}
\begin{center}
\includegraphics[height=3.0in,width=3.0in]{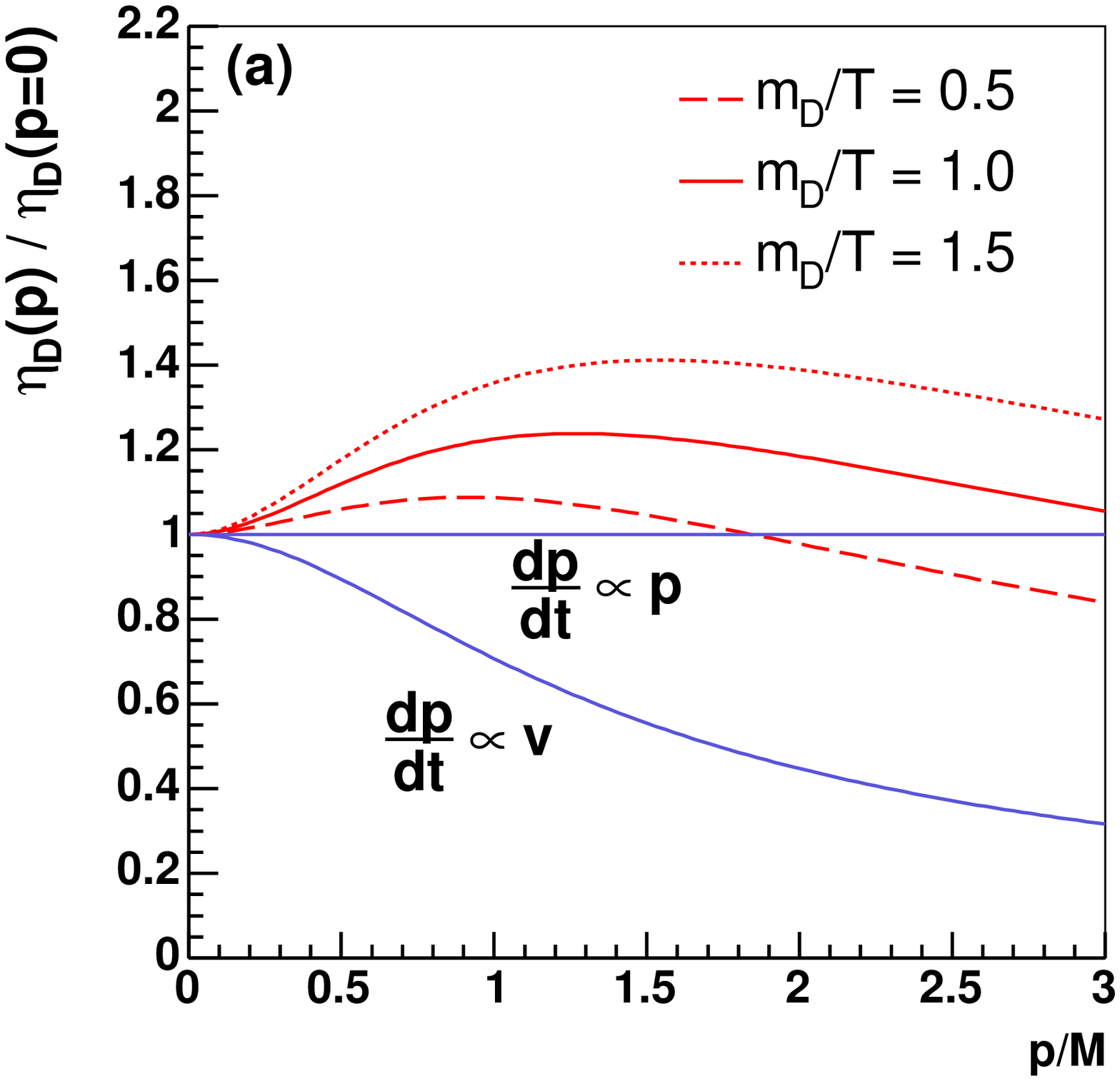}
\includegraphics[height=3.0in,width=3.0in]{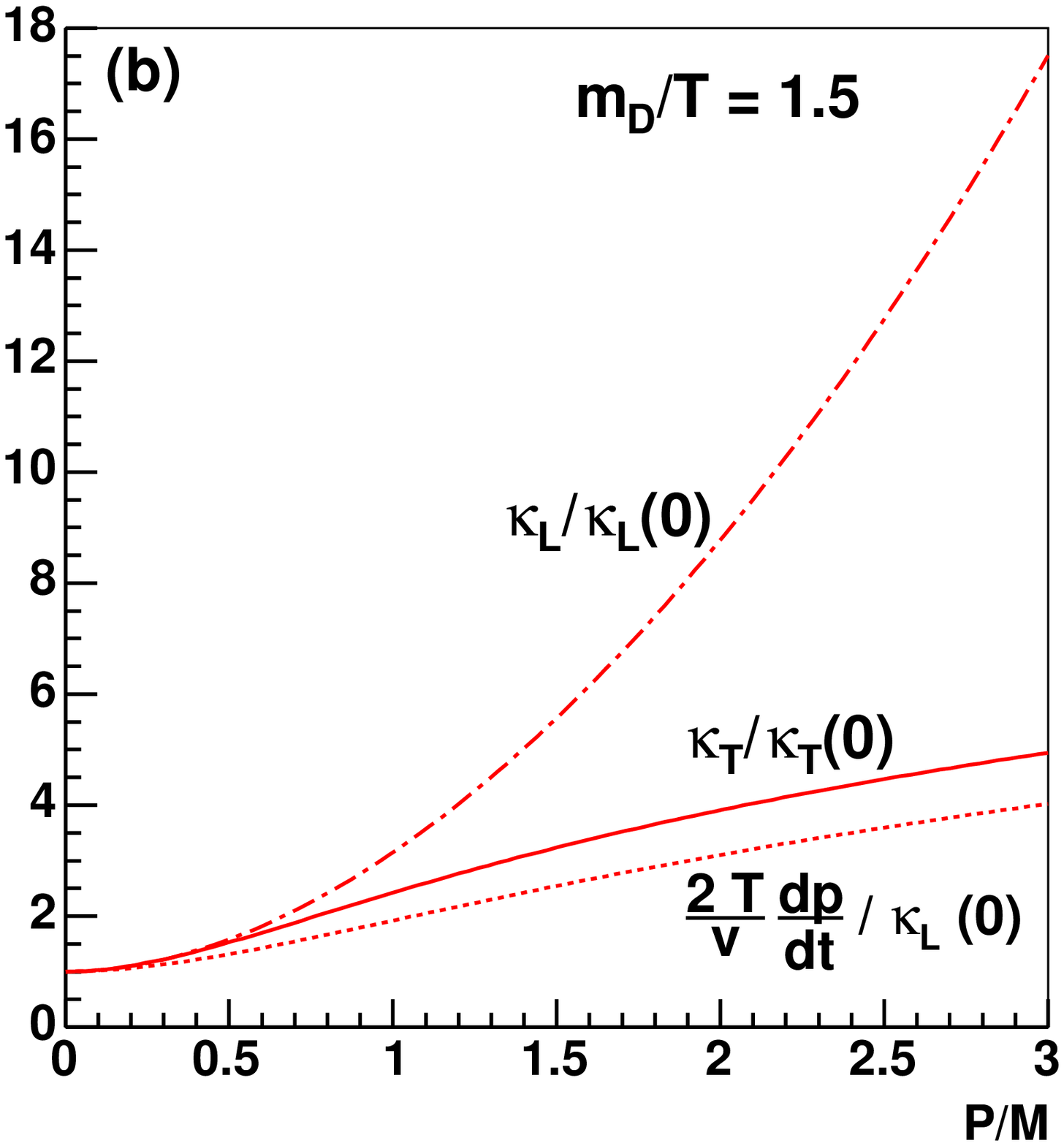}
\caption{ (Color online)
(a) The drag coefficient as a function of the heavy quark momentum,
$\frac{dp}{dt}=\eta_D(p) p$.
(b) The transverse ($\kappa_T(p)$) and longitudinal ($\kappa_L(p)$)
momentum diffusion coefficients
as a function of heavy quark momentum for $m_{D}/T=1.5$ . For comparison
we also show the longitudinal momentum diffusion coefficient
in a Boltzmann-Langevin approach ($ \frac{2 T}{v}\, \frac{dp}{dt}$).  }
\label{transportc}
\end{center}
\end{figure}

As for the  diffusion coefficient, these expressions are
only valid when the Debye mass is very small compared to the
temperature. To extrapolate to finite $\mD/T$ we return
to the original expressions and numerically
perform the phase-space integrals to determine the transport
coefficients.
The phase space integrals can be reduced to three dimensional
integrals as is detailed in the appendix. The resulting transport
coefficients and their dependence on the heavy quark momentum are
illustrated in Fig.~\ref{transportc} and are discussed below.

The momentum dependence of the drag coefficient
for various values of $m_D/T$ is shown
in Fig.~\ref{transportc}(a).  As above, we emphasize
that these results are strictly valid only
when the Debye mass is small and therefore the
different curves illustrate the uncertainty in the
calculation. The drag coefficient $\eta_D(p)$
has units $(\mbox{time})^{-1}$ and sets the
equilibration rate.
Since the drag coefficient at zero momentum is related to
the diffusion coefficient which has
already been discussed,
we have divided by the drag coefficient at zero momentum to isolate
the momentum dependence.
We see that the equilibration rate $\eta_D(p)$ does not decrease significantly
with momentum. Only when the momentum is larger
than $3 M$ does the drag coefficient decrease.
For comparison we have illustrated the expected form of the drag
coefficient for an extreme  model which is studied
later. In this model, we  have $\frac{dp}{dt} \propto v$ and
therefore the equilibration time is inversely proportional
to the energy of the heavy quark, $\eta_{D} \propto 1/E$.

Next we report on the transverse and longitudinal fluctuations
for the heavy quark in Fig.~\ref{transportc}(b). We see that both the
transverse and longitudinal fluctuations ($\kappa_T(p)/\kappa_T(0)$ and
$\kappa_L(p)/\kappa_L(0)$) rise with
the momentum of the heavy quark.
It is instructive to compare the longitudinal fluctuations  derived from
the full Boltzmann equation to the longitudinal fluctuations expected in
a  Boltzmann-Langevin approach described in the
next section. In the Boltzmann-Langevin approach the
 longitudinal fluctuations are directly related to the drag
through the relation
\[
         \frac{d}{dt} \, \left\langle (\Delta p_z)^2 \right\rangle = \frac{2 T}{v}\, \frac{dp}{dt} \; .
\]
We see that the Boltzmann-Langevin prediction for the
longitudinal fluctuations ($ \frac{2 T}{v}\, \frac{dp}{dt}$) underestimates
the fluctuations of the full Boltzmann equation ($\kappa_L$).  At large
momenta the underestimate can be a factor of 2 to 4.

\section{The Boltzmann Fokker-Planck Equation}

The discussion in the previous sections (especially the first paragraph of \Sect{eloss-sect})
suggests a relativistic generalization of the non-relativistic
Langevin equations. To this end, we
write a stochastic equation of motion for the
heavy quark in the rest frame of the medium
\begin{eqnarray}
\label{langevin}
    \frac{ dp_{L} }{ dt } &=& -\eta_D(p) p^{i} + \xi_L \; , \\
    \frac{ dp_{T} }{ dt}  &=& \xi_T  \; .
\end{eqnarray}
Here $dp_{L}$ and $dp_T$ are the momentum increments parallel and
transverse to the direction of the heavy quark respectively.
$\xi_L$ and $\xi_T$
are random momentum kicks in the longitudinal and transverse
directions which satisfy
\begin{eqnarray}
\langle \xi_L^{i}(t)\, \xi_L^j(t') \rangle &=&
\kappa_L(p) \,
\hat{p}^i \hat{p}^j
	\delta(t-t') \; ,\\
\langle \xi_T^{i}(t)\, \xi_T^j(t') \rangle &=&
\kappa_T(p)
\, (\delta^{ij} - \hat{p}^i \hat{p}^j)
	\delta(t-t')  \; , \\
\langle \xi_T^{i}(t)\, \xi_L^j(t') \rangle &=&  0 \; .
\end{eqnarray}
Thus $\eta_{D}(p) p$  is the momentum loss per unit
$\Delta t$ and  $\kappa_T(p)$ and $\kappa_L(p)$  are the
variances of the transverse and longitudinal momentum transfers
per unit $\Delta t$. From the discussion in \Sect{eloss-sect},  the precise momentum
where $\eta_D(p)$ should be evaluated is known only up
to corrections of order $T/M$.

The Langevin equation is ambiguous  until it is discretized.
If time is divided into discrete steps of $\Delta t$  and the
the momenta at discrete times are labeled  $\p^{0}, \p^{1}, ...,\p^{n}$,
then the Ito discretization of the Langevin equation is
\[
    (p^{n+1})^i - (p^{n})^{i}  =  a^i_{\ito}(\p^{n})\,\Delta t + {\xi^i}(\p^{n}) \Delta t  \; ,
\]
where $\xi^i(\p^{n})$ is drawn from the a Gaussian distribution
such that
\[
     \left< \xi^i(\p^n) \xi^{j}(\p^m) \right > =
     b^{ij}(\p^{n})\frac{\delta^{mn}}{\Delta t} \; ,
\]
and we have defined to coefficients $a^{i}(\p)$ and $b^{ij}(\p)$
as
\begin{eqnarray}
a^{i}_{\ito}(\p) &\equiv& -\eta_D(p) p^{i} \; ,\\
b^{ij}(\p) &\equiv& \kappa_L(p)\,\hat{p}^i \hat{p}^j +
            \kappa_T(p) \,(\delta^{ij} - \hat{p}^i\hat{p}^j ) \; .
\end{eqnarray}

With this choice of discretization, the Langevin equation
is equivalent to a Fokker-Planck equation,
\st
  \label{fpv}
  \frac{ \partial P  }{\partial t }  +
  \frac{ \partial } { \partial p^{i} }
  ( a^{i}_{\ito}(\p) P ) -
  \frac{1}{2}
  \frac{ \partial^2 } { \partial p^{i}\, \partial p^{j} }
  ( b^{ij}(\p) P )  =  0  \; .
\stp
The most instructive way to show the equivalence
between the Fokker-Plank and Langevin approach
is to recognize that the Fokker-Planck equation
is a Euclidean Schr\"odinger equation and therefore has
a phase space path integral representation for the transition
probability $P(\p,t|\p_{0},t_0)$.
(Here the canonical momenta and canonical coordinates are
$\Pi=\frac{\partial}{\partial \p}$ and $Q = \p$, respectively.)
Similarly it is easy to write down a path integral expression
for the transition probability from the Langevin equations.
The two path integrals are the same after integrating
over canonical momenta \cite{Arnold:1999uz}.

Another equally valid discretization
is the Stratonovich discretization of the Langevin
equation
\[
    (p^{n+1})^i - (p^{n})^i  = a^i_{\strat}(\bar{\p})\, \Delta t + {\xi^i}(\bar{\p}) {\Delta t}  \; .
\]
where $\bar{\p}$ = $(\p^{n+1} + \p^n)/2$.
This discretization will lead to a Fokker-Planck equation of the a
slightly different form
\[
 \label{fpv2}
  \frac{\partial P}{\partial t} +
  \frac{ \partial } {\partial p^i } ( a^{i}_{\strat}(\p) P )
  - \frac{1}{2}
  \frac{ \partial } {\partial p^i } (
  b^{ij}(\p)
  \frac{ \partial P} { \partial p^j} ) = 0 \; .
\]
Clearly the two forms of the Fokker-Planck equation will give the same answer if
\st
\label{conversion}
   a^{i}_{\strat}(p) = a^i_{\ito}(\p) - \frac{1}{2} \frac{\partial b^{ij}(\p)}{\partial p_j} \; .
\stp

The resolution of this ambiguity was clarified by Arnold \cite{Arnold:1999uz}.
The correct procedure is to adjust the drag coefficient $a^{i}$ so that
the Fokker-Planck equation approaches equilibrium.
We will discuss the Ito case here.
Substituting the equilibrium distribution $P \propto e^{-E_p/T}$ into \Eq{fpv}
and demanding that the r.h.s. return zero we obtain a relation between
$a^{i}_{\ito}(\p)$ and $b^{ij}(\p)$. This relation, in terms of $\eta_D$,
$\kappa_L$, and  $\kappa_T$ is
\begin{eqnarray}
\label{etadv}
    \eta_D^{(0)}  (v) &\equiv& \frac{\kappa_L(v)}{2 T E}  \; ,\nonumber \\
    \eta_D^{\ito} (v) &=& \eta_{D}^{(0)} - \frac{1}{ E^2}
    \left[
         (1-v^2)
         \frac{\partial}{\partial v^2} (\kappa_L(v))
         + \frac{d}{2} \frac{(\kappa_L(v) - \kappa_T)v))} {v^2}
    \right]     \; .
\end{eqnarray}
Here $d=3$ is the number of dimensions and  $v$ is the velocity
of the heavy quark. The derivative term
in square brackets is smaller by a factor of $T/E$ than the first
term and serves to renormalize the drag coefficient \cite{Arnold:1999uz}.
The derivative corrections to the drag coefficient $a^{i}$ reflect
the ambiguity in the momentum  at which $a^{i}$ is to be evaluated.
Similarly, in the Stratonovich case we  have
$\eta_D^{\strat} \approx \eta_{D}^{(0)} + O(T/E)$. The $O(T/E)$ term
can be deduced from the relation between the Stratonovich and Ito
discretizations, \Eq{conversion}.

In the previous section we computed in kinetic theory the
drag coefficient
to leading order in $T/E$, $\eta_{D}^{(0)}$. We
also computed the longitudinal fluctuations $\kappa_L$
to leading order in $T/E$.
For the Fokker-Planck approach to be strictly valid
the relation $\eta_{D}^{(0)}(p) = \kappa_L(p)/(2\,T\,E)$ must be satisfied;
otherwise the stochastic process will not approach equilibrium.
This is equivalent to the requirement that
\begin{equation}
\label{fdr}
    \frac{d}{dt} \llangle (\Delta p_z)^2 \rrangle = \frac{2\,T}{v} \frac{dp}{dt}  \; .
\end{equation}
{}From kinetic theory we have  the following equations
for the drag and fluctuations:
\begin{eqnarray*}
    \frac{dp}{dt}  &=&
\frac{1}{2v} \int_{k,q} |\M|^2 \,q^0 \,
\left\{ f[k] (1\pm f[k{-}q^0]) - f[k{-}q^0] (1\pm f[k]) \right\} \; , \\
    \frac{d}{dt} \llangle (\Delta p_z)^2 \rrangle &=&
 \int_{k,q} |\M|^2 \, q_z^2 \,
	\left\{ f[k] (1\pm f[k{-}q^0]) \right\} \; .
\end{eqnarray*}
In general, these two equations are not related to each other by the
fluctuation dissipation relation, \Eq{fdr}.
However, if the transfer energy is small, $\omega \ll T$, we can expand the
thermal distribution functions in the drag equation to the leading non-trivial
order in $\omega$ to find
\begin{eqnarray*}
    \frac{dp}{dt}  &=&
\frac{1}{2 v T} \int_{k,q} |\M|^2 \,\omega^2 \,
\left\{ f[k] (1\pm f[k]) \right\} \; ,  \\
    \frac{d}{dt} \left\langle (\Delta p_z)^2 \right\rangle &=&
 \int_{k,q} |\M|^2 \, q_z^2 \, \left\{ f[k] (1\pm f[k])  \right\} \; .
\end{eqnarray*}
Upon using the simple kinematic formula $\omega = v\, q_z$ given
in Appendix \ref{appendixA}, we find the required relation \Eq{fdr}. Thus
only when the transfer energy is small compared to the temperature
is the Langevin model strictly valid.

The transfer energy is small only in two limiting cases. In the first
case  the heavy quark is non-relativistic. In this limit  the
energy transfer $\omega$ is less than the velocity times the momentum transfer,
$\omega \leq v\,q$  and therefore is small compared to to the temperature. If
the quark is relativistic, however, $\omega$ is only small if the momentum
transfer $q$ is small compared to $T$. This is true only to leading-log of
$T/m_D$. Thus we see that the leading-log transport coefficients
(\Eq{dpdtll} and \Eq{eq:leadinglog})  satisfy the fluctuation
dissipation relation, \Eq{fdr}.

This analysis indicates that  an amalgamation of kinetic theory and the
Langevin approach is correct to leading order. Given the two
scales $m_D$ and $T$,  introduce $q^{*}$ between $m_D$ and $T$, say
$q^{*} \sim \sqrt{m_D T}$. Then treat all collisions with
large momentum transfer $q > q^{*} \gg m_D$ with ordinary unscreened
kinetic theory. The kinetics of the hard collisions depends
logarithmically on $T/q^{*}$. All collisions with momentum transfer $q<q^{*} \ll T$ can
be subsumed into a Langevin process with prescribed transport coefficients
that depend logarithmically on $q^{*}/m_{D}$. The dependence on
$q^{*}$ cancels when both the Langevin process and hard collisions are
included. This procedure is
only useful when $m_D$ is really much smaller than $T$ and we will not
adopt it.

\section{A Langevin Model For Heavy Ion Collisions}
\label{Bjorken}

We have seen that, excepting non-relativistic quarks,
the Langevin/Fokker-Planck approach is a valid description
of the kinetics of heavy quarks only to leading logarithm in $T/m_D$.
Since the coupling is not particularly small, this result
says that the full kinetic theory should be used to
find the evolution of the heavy quark spectrum.

However, the Langevin model is an appealingly simple framework
for studying the thermalization of heavy quarks in a heavy ion collision.
The Langevin model requires two inputs, the drag coefficient $\eta_D^{(0)}(v)$ and the transverse momentum fluctuations
$\kappa_T(v)$. The longitudinal fluctuations  are related to $\eta_D^{(0)}(v)$
by the fluctuation dissipation relation $\eta_D^{(0)}(v) = \kappa_L(v)/(2 T E)$.
(Finally the  the drag coefficient can be tweaked as in \Eq{etadv} by terms
suppressed by $T/E$ so that the Langevin process approaches equilibrium. )
As seen in Fig.~\ref{transportc}(b), the longitudinal fluctuations in the Langevin model ($\frac{2T}{v} \frac{dp}{dt}$) are generally
smaller than the corresponding fluctuations ($\kappa_L(p)$) in the full kinetic
theory.
The Langevin process will underestimate the longitudinal diffusion at high
momentum by a factor of $2$ to $4$.

We have employed two models for the drag and transverse momentum diffusion
coefficients.
The first model is based on LO thermal QCD computation of the
drag and transverse momentum diffusion coefficients. The procedure to set these
coefficients is the following. First we set $m_D/T = 1.5$ and then
use the results of \Sect{eloss-sect} to determine the momentum dependence of the coefficients,
$\eta_D(p)/\eta_D(0)$ and $\kappa_T(p)/\kappa_T(0)$. Then $\eta_D(0)$ and
$\kappa_T(0) =\kappa_L(0)$ are fixed by specifying the coefficient $D$ and using
the relation
\[
      \eta_{D}(0)^{-1} = \frac{M}{T} D = \frac{2 T M}{\kappa_{L}(0)}\; .
\]
This procedure is equivalent to simply adjusting
$\alphas$ while leaving $m_D/T$ fixed in the equations.
This model is referred to as LO QCD below.

The second model is an extreme limit. In this case we set
$\kappa_L$ equal to a constant, which is equivalent to setting
\[
     \frac{dp}{dt}  \propto {v} \; .
\]
The
transverse and longitudinal diffusion coefficients
are then set equal to each other, $\kappa_T(v)=\kappa_L(v)$.
The precise value of $\kappa_L(v)$ is specified by setting
the diffusion coefficient as in the previous model.

We have also considered a third extreme limit. In this
case we set $\eta_D(p)$ equal to a constant, which is equivalent
to setting
\[
    \frac{dp}{dt} \propto {p} \; .
\]
We have found that the results of this model are quite
similar to the leading order model LO QCD as could
have been intuited from \Fig{transportc}(a). We will not
discuss this model further.

\subsection{Solution to the Fokker-Planck Equation for a Bjorken Expansion}

Before considering this Langevin model for the drag and
diffusion in detail, let us estimate the
effects of thermalization on the spectrum of
non-relativistic heavy quarks in a medium expanding in a boost
invariant fashion.
In this section we assume that the transport
coefficients are evaluated at $v=0$ where
they are all related by constants, {\it i.e.\ }
$\kappa = \kappa_L(0) = \kappa_T(0) = (2 T M) \eta_D(0)$.  In
the non-relativistic limit the Langevin approach is strictly valid.

The Langevin update rule in the local rest frame is
\begin{eqnarray}
     \Delta \x &=& \frac{\p}{M} \Delta t \; , \\
     \Delta \p &=& (-\eta_{D} \p  + {\bf \xi} )\, \Delta t  \; ,
\end{eqnarray}
where ${\bf \xi}$ is drawn from a Gaussian probability distribution
of width  $\left\langle \xi^{i}\xi^{j}\right\rangle = \kappa/{\Delta t} \delta^{ij}$.
The Boltzmann-Fokker Planck Equation (BFPE) for this
update rule is
\begin{equation}
\frac{ \partial P }{\partial t} + \frac{p^{i}}{M} \frac{ \partial P }
{\partial x^{i} } =
   \left[ \frac{ \partial} {\partial p^i}  \eta_D  p^i   +
   \frac{ \partial^2 } {\partial p^2 }   M T \;\eta_D  \right] P(\x,\p,t)  \; ,
\label{bfpe}
\end{equation}
where $P(\x,\p,t)$ denotes the probability to
observe a quark with position $\x$ and momentum $\p$ at time $t$.  Our
goal in the following paragraphs is to solve this partial differential
equation for the evolution of the probability distribution for the
simple case of a Bjorken expansion.

For a Bjorken expansion \cite{Bjorken:1982qr}, the probability
distribution is independent of the transverse coordinates and invariant
under boosts in the $z$ direction. This allows the BFPE
to be simplified \cite{Baym:1984np}, and the Green  function of the
resulting partial differential equation is
obtained in Appendix \ref{appendixBj}.
To obtain the final spectrum at time $t$
we need to convolve this Green function $P(\p,t|\p_0,t_0)$ with
the initial spectrum at time $t_0$,
\begin{eqnarray}
\label{convolve}
 \left. \frac{dN}{d^3p\, d\eta} \right|_{\eta=0,t}
  =  \int d^3p_0 P(\p,t | \p_0,t_0)
           \, \left. \frac{dN}{d^3p_0\, d\eta} \right|_{\eta=0,t_0} \; .
\end{eqnarray}
In this restricted sense of \Eq{convolve}, $P(\p,t|\p_0,t_0)$ is the
probability that the particle
with momentum $\p_{0}$ at time $t^{0}$ evolves to
a particle with momentum $\p$  at some time $t$. $P(\p,t|\p_0,t_0)$ is
given by
\begin{eqnarray}
P(\p, t | \p_0, t_0)
 &=&
 \frac{ 1}{\sqrt{ 2 \pi M T^{\perp}(t) }  }
 \exp\left[ - \frac{ (p_x - p_x^{0} e^{-\chi(t)})^2 }{2 M T^{\perp}(t) }
 \right]  \nonumber \\
 &\times&
 \frac{ 1}{\sqrt{ 2 \pi M T^{\perp}(t) }  }
  \exp\left[ - \frac{ (p_y - p_y^{0} e^{-\chi(t)})^2 }{2 M T^{\perp}(t)
 } \right] \nonumber \\
  &\times &
 \frac{ 1}{\sqrt{ 2 \pi M T^{z}(t) }  }
 \exp\left[ - \frac{ (p_z - \left(\frac{t_0}{t}\right)\; p_z^{0} \; e^{-\chi(t)})^2 }{2 M T^{z}(t) } \right] \; ,
\end{eqnarray}
where we have defined the transverse and longitudinal effective temperatures,
\begin{eqnarray}
\label{teff}
   \chi(t) &=&  \int_{t_0}^{t}dt' \;\eta_D(t') \; , \\
   T^\perp(t) &=&   2\int_{t_0}^{t}dt' \;\eta_D(t')  \; T(t') \;
  e^{2 \chi(t')  - 2\chi(t) } \; , \\
   T^{z}(t) &=&   2\int_{t_0}^{t}dt' \;\eta_D(t') \;  T(t')
   \left( \frac{t'}{t} \right)^2
  e^{2 \chi(t')  - 2\chi(t) } \; .
\end{eqnarray}

\begin{figure}
\begin{center}
\includegraphics[height=3.5in,width=3.5in, angle=-90]{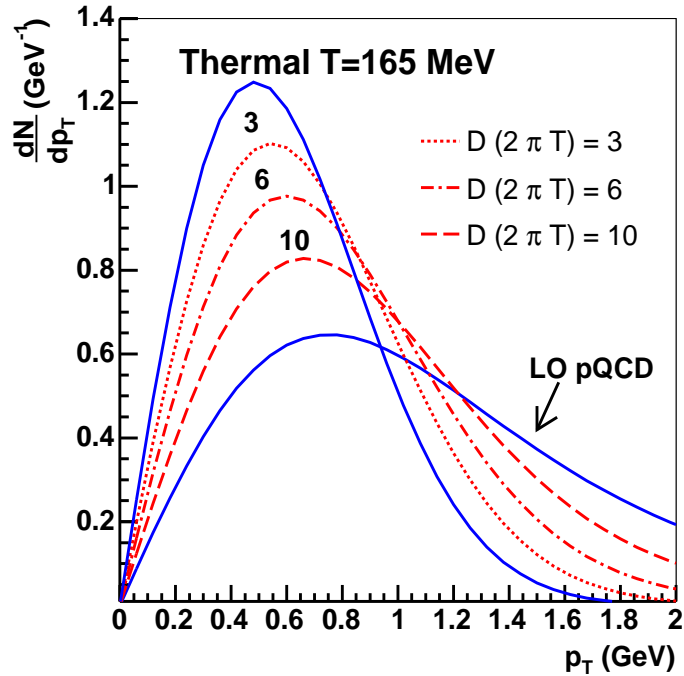}
\caption{ (Color online)
The transverse momentum spectrum of charm quarks at time $t_f=6\,\mbox{fm}$ for
a Bjorken expansion with $\tau_0=1\,\mbox{fm}$ and $T_0=300\,\mbox{MeV}$ and
$T(t_f)=165\,\mbox{MeV}$.
The initial transverse momentum spectrum is given by leading order
perturbation theory (LO pQCD). }
\label{distribution}
\end{center}
\end{figure}

Now we will estimate how the initial spectrum
of heavy charm quarks is changed
by the interactions.
For simplicity consider a Bjorken expansion with
$T(t_0) = 300\,\mbox{MeV}$, $t_0=1\,\mbox{fm}$, $M=1.4\,\mbox{GeV}$.
For an ideal Bjorken expansion, the temperature  and
drag coefficients follow \cite{Bjorken:1982qr}
\begin{eqnarray}
          T(t) &=&  T(t_0) \left( \frac{t}{t_0} \right)^{-\frac{1}{3} } \; ,\\
     \eta_D(t) &=&  \eta_D(t_0) \left( \frac{t}{t_0} \right)^{
     -\frac{2}{3} } \; .
\end{eqnarray}
$\eta_D(t_0)$ is adjusted according to \Eq{eq:D} to give
a specified diffusion coefficient.
We will compute the spectrum of charm quarks at a  final time
$t_f = 6\,\mbox{fm}$. The equilibrium temperature at this time is
$T(t_f)=165\,\mbox{MeV}$.
For the initial spectrum of heavy quarks we will
take the transverse momentum spectrum from a
fit to leading order parton model calculations which are described
in the next section,
\begin{eqnarray}
    \frac{ dN }
    {d\eta\, dy d^2p_{T} } &\propto&
    \delta(y-\eta) \frac{1}{(p_{T}^2 + \Lambda^2)^{\alpha} } \; ,
\end{eqnarray}
with $\alpha=3.52$ and $\Lambda=1.85\,\mbox{GeV}$. The initial
spectrum is shown in Fig.~\ref{distribution} and is labeled
as LO pQCD.
Then we convolve this initial condition with the  Green function
as in \Eq{convolve} to determine the final spectrum.

The final transverse momentum spectrum is shown as a function
of the diffusion coefficient in Fig.~\ref{distribution}.
We observe that the charm spectrum approaches the
thermal spectrum only when the diffusion coefficient
is less than $3/(2\pi T)$. This is small
relative to the estimates made in the previous section. Further,
such a small diffusion coefficient
implies a substantial suppression of the spectrum at
large transverse momentum. This basic observation
will be quantitatively confirmed when we include radial
flow in the next section.  As studied in
Appendix\,~\ref{appendixBj}, for large diffusion coefficients
interactions simply smear the original spectrum (LO pQCD)
with a Gaussian. For small diffusion coefficients the spectrum
is close to the thermal spectrum ($T=165\,\mbox{MeV}$) up to small viscous corrections.

\subsection{Elliptic Flow and Suppression of Charm Quarks}

Next we will calculate how the flow of an underlying medium
influences the spectrum of heavy quarks. If the relaxation
time $\eta_D^{-1}$ is less than the expansion rate of
the medium, then the heavy quark will follow the medium. If
$\eta_D^{-1}$ is greater than the expansion rate, the
heavy quark will not follow the medium and the resulting
elliptic flow will be small. Of course, the relaxation time
depends on the momentum of the heavy quark. The goal
here is to determine the largest possible elliptic flow
for a given value of the diffusion coefficient.

To this end we have placed heavy quarks
into a hydrodynamic simulation of the heavy ion collision.
In the local rest frame of the medium, the heavy quark follows
the Langevin equations. Further discussion of
Lorentz invariance and numerical implementation is
given below.

The hydrodynamic simulation is a $2+1$ boost invariant
hydrodynamic model with an ideal gas equation
of state $p = \frac{1}{3}\,e$. The temperature is
related to the energy density with the $N_f=3$ ideal
QGP equation of state.  We have chosen this
extreme equation of state because the resulting radial
and elliptic flow are too large relative to data on
light hadron production. Thus,
this equation of state will estimate the largest elliptic
flow possible for a given diffusion coefficient.

Aside from the equation of state, the hydrodynamic model is based upon
References \cite{Teaney:2001av}.
At an initial time $\tau_0=1.0\,\mbox{fm}$,
the entropy is distributed in the transverse plane
according to the distribution of wounded nucleons
for a Au-Au collision with an impact parameter of $b=6.5\,\mbox{fm}$.
Then one parameter $s_0=14\,\mbox{fm}^{-2}$, which
is the entropy per unit rapidity per wounded nucleon per area,
is  adjusted to set the initial temperature and total particle yield.
The value  $s_0=14\,\mbox{fm}^{-2}$ closely corresponds to the
results of full hydrodynamic simulations \cite{Teaney:2001av,Kolb:2000fh,Huovinen:2001cy} and  corresponds to a maximum initial
temperature of $T_{0} = 265\,\mbox{MeV}$.

At the initial Bjorken time $\tau_0$, the position and momentum
distributions of
the heavy quarks are estimated from leading order parton model calculations. To
this end, we have distributed the heavy quarks (about a million or so per run)
in the transverse plane according to the distribution of binary collisions.  In
the longitudinal direction the heavy quarks are distributed uniformly in a
large space-time rapidity window $\eta = -20...20$.  Periodic boundary
conditions are applied in the $\eta$ direction.  The initial momentum
distribution is drawn from
\begin{eqnarray}
    \frac{dN}{dy\,d\eta\, d^2p_T} \propto \delta(y-\eta)
    \frac{1}{(p_T^2 + \Lambda^2)^{\alpha}  } \; ,
\end{eqnarray}
where $\alpha = 3.5$ and $\Lambda=1.849\,\mbox{GeV}$. This parameterization
is a fit to leading order parton model calculations
which were performed with the marvelous CompHEP package ~\cite{CompHEP}.
Specifically, charm production at mid-rapidity was computed for proton-proton
collisions with $\sqrt{s}=200\, \mbox{GeV}$.  The charm mass was set to
$1.4$ GeV and the strong coupling constant was evaluated at a scale
$4m_{T}^2=\hat{s}\left[1 - (-t + m_c^2/\hat{s})^2\right]$. Similarly the CETQ4m
parton distributions were evaluated at a scale $4 m_{T}^2$.  The resulting
momentum distribution is comparable to the results of Pythia
calculations \cite{Pythia}.

A few remarks about Lorentz invariance and numerical implementation
are necessary.  Consider a heavy quark with position and momentum
4-vectors $(x')^{\mu}$ and $(p')^{\mu}$ in
a medium with four-velocity $u^{\mu}(x')$ in the computational frame.
Given an infinitesimal time interval $\Delta t'$,  we
can calculate $\Delta \x'$ in the computational frame, {\it i.e.\ }
$\Delta \x' = \frac{\p'}{E'} \Delta t'$. We therefore know the
4-vector $(\Delta x')^{\mu}$ and we can compute $(\Delta x)^{\mu}$
in the rest frame of the medium.
In particular, we know the time interval in the rest frame $\Delta t$.
In the rest frame, we can update the momentum using the Langevin rule.
Now since the quark is on mass shell, the four momentum is known in the rest
frame and can be boosted back to the computational frame. Generally there will
be numerical artifacts which are not Lorentz invariant of order $\sqrt{\Delta
t/L}$, where $L$ is some typical time/length scale of change in the
computational frame. Because of this dependence on $\Delta t$, it is important
to check that Lorentz invariance is respected. Several test runs were performed
in different frames to verify that numerical artifacts are under control at the
$1.0\%$ level when the $\gamma$ factor $u^{0}$ is less than $15$.  The
$\sqrt{\Delta t}$ error can presumably be avoided by employing a higher order
algorithm for stochastic differential equations \cite{hosde}. But, given the
complexity of the intermediate boosts, the lowest order algorithm was adopted.

In summary, we place heavy quarks with a reasonable initial transverse momentum
distribution in a hydrodynamic simulation of heavy ion collisions. Then the
heavy quarks are evolved according the the Langevin update rule in the local
rest frame. We have simulated a solidly mid-central collision,
$b=6.5\,\mbox{fm}$.

\begin{figure}
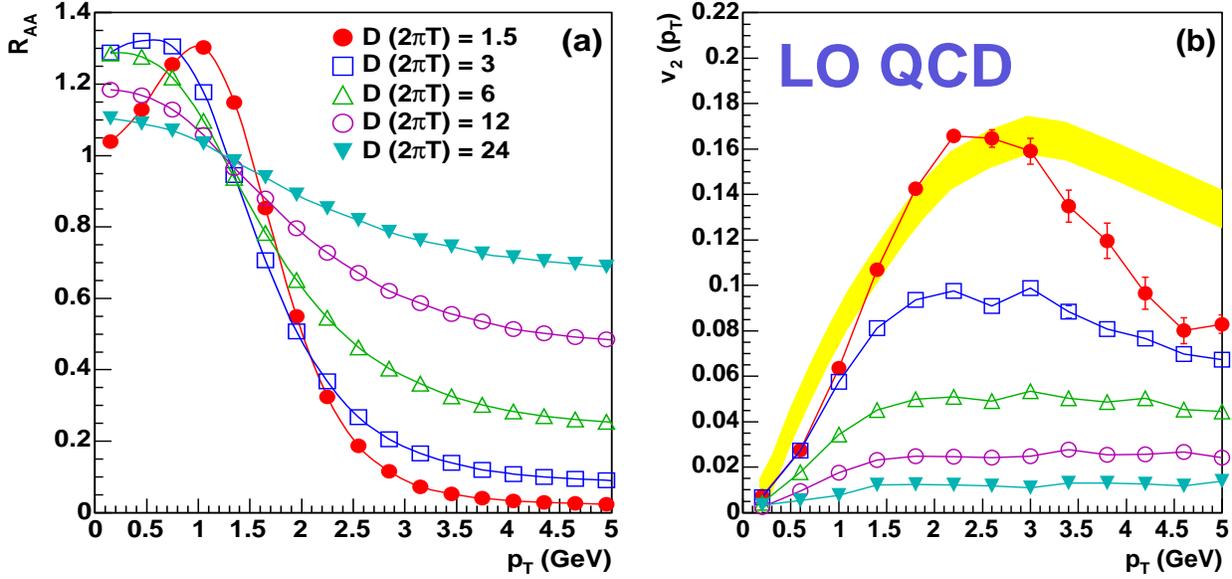

\begin{center}
\includegraphics[height=3.2in,width=3.0in, angle=-90]{qcd15_rpa.epsi}
\hspace{0.1in}
\includegraphics[height=3.0in,width=3.0in, angle=-90,clip=false]{qcd15_v2.epsi}
\caption{ (Color online)
(a) The nuclear modification factor $R_{AA}$ for charm quarks for
representative values of the diffusion coefficient. (b) $v_2(p_T)$ for
charm quarks for the same set of diffusion coefficients given in the
legend in (a). In perturbation
theory, $D \times (2\pi T) \approx 6\,(0.5/\alphas)^2$. The model
for the drag and fluctuation coefficients is referred to as LO QCD in
the text. The band estimates
the light hadron elliptic flow for impact parameter $b=6.5\,\mbox{fm}$ using
STAR data \protect\cite{Adams:2004bi}.}
\label{RAA}
\end{center}
\end{figure}

There are at least two items of experimental
interest in the heavy ion program. The first is the suppression factor
$R_{AA}$, which experimentally
is the ratio of the proton-proton $D$ meson spectrum
to the Au-Au $D$ meson spectrum. We will estimate $R_{AA}$ at mid-rapidity
using our ``input'' and ``output'' charm quark spectrum,
\[
     R_{AA} = \frac{ \left( \frac{dN}{dp_T} \right)_{\mbox{output}} }{
     \left( \frac{dN}{dp_T} \right)_{\mbox{input}} } \; .
\]
The ``input'' spectrum  has already been described.
Given the Langevin nature of the force, the spectrum never
completely stops changing. However, the drag and fluctuations are
proportional to powers of the temperature, and therefore their influence
decreases at late times, as the temperature drops. We have
found that the spectrum is nearly frozen after $\tau = 9\,\mbox{fm}$.
The ``output'' spectrum is really the spectrum at $\tau = 15\,\mbox{fm}$.
Integrating further would have only a negligible effect.  A large unknown
in the comparison to $D$ meson data is how hadronization will affect
the final spectrum. In the truly heavy quark limit, hadronization would
not affect the spectrum significantly. However, the charm quark is
not particularly heavy, and therefore various models of hadronization
will give different answers.
Coalescence and independent fragmentation are two extreme models which
have been studied recently \cite{Lin:2003jy,Greco:2003vf}.
We will ignore the important effects of hadronization in this work.
Therefore, a direct comparison of our $R_{AA}$ to the experimental $R_{AA}$
is certainly misguided.  The $R_{AA}$ factor is illustrated in
Fig.~\ref{RAA}(a).  This figure is discussed below.

A second item of experimental interest is  elliptic flow.
For simplicity we will consider only mid-rapidity.
Elliptic flow is quantified with $v_{2}(p_{T})$,
\[
   v_{2}(p_T) = \left\langle \cos(2\phi) \right\rangle_{p_T}
              = \frac{
                     \int d\phi \, \left. \frac{dN}{dy dp_T d\phi}\right|_{y=0} \,
                     \cos(2\phi)
                     }{
                     \int d\phi \, \left. \frac{dN}{dy dp_T d\phi}\right|_{y=0}
                     } \; .
\]
Here the angle $\phi$ is measured with respect to the reaction
plane.  The experimental determination of the reaction plane is
discussed in \cite{Borghini:2000sa,Bhalerao:2003yq}. Again, we will
estimate the elliptic flow of $D$ mesons with our charm quark spectrum
and ignore potential modifications due to hadronization.
$v_2(p_T)$ is illustrated in Fig.~\ref{RAA}(b).

\begin{figure}
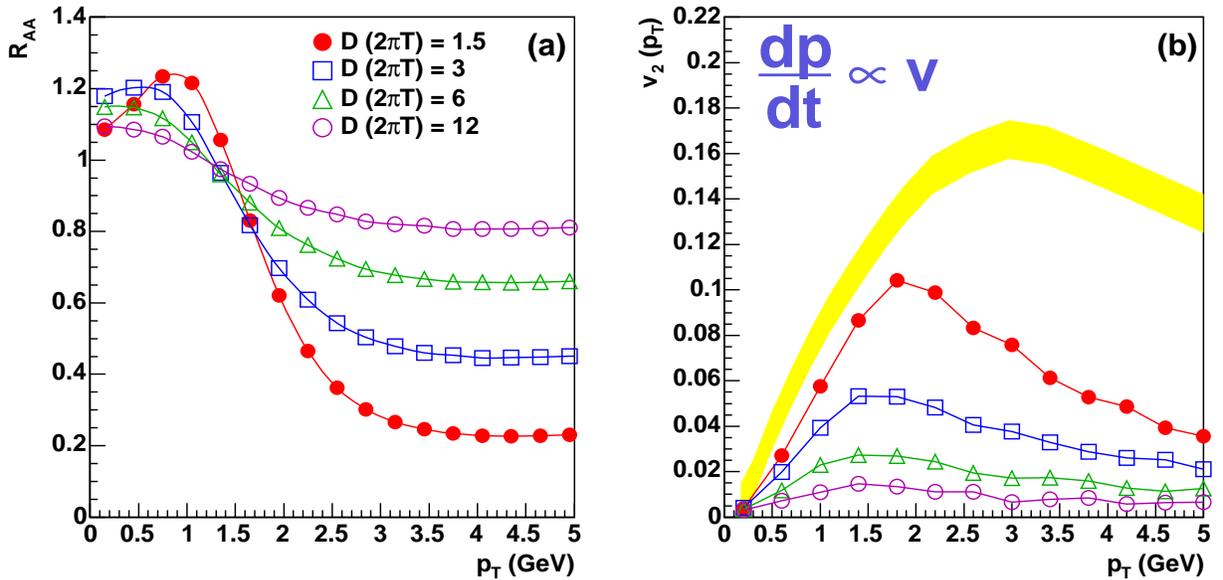

\begin{center}
\includegraphics[height=3.0in,width=3.0in, angle=-90]{constk15_rpa.epsi}
\hspace{0.2in}
\includegraphics[height=3.0in,width=3.0in, angle=-90]{constk15_v2.epsi}
\caption{ (Color online)
(a) The charm quark nuclear modification factor $R_{AA}$  and  (b) elliptic flow for representative values of the diffusion coefficient given in the
legend.
In this model, the drag is proportional to the velocity,
$\frac{dp}{dt}\propto v$.  For further explanation see Fig.~\ref{RAA}.}
\label{RAA2}
\end{center}
\end{figure}

To illustrate the model dependence, we have also calculated $R_{AA}$ and
$v_2(p_T)$ using two extreme models for the drag and fluctuations.
As already elaborated, in the first model the drag is proportional to
the velocity,
$\frac{dp}{dt} \propto v$.  The methods and caveats of extracting $R_{AA}$
and $v_2(p_T)$ from the simulation are the same.
The results are shown in Fig.~\ref{RAA2}(a) and (b).  In the second
model, the drag is proportional to the momentum, $\frac{dp}{dt} \propto p$.
However, the results are not too different from the LO QCD model
and therefore the results are not shown.

%
\subsection{Comparison with Boltzmann Simulations}
\label{comparison}

It is useful to compare our Boltzmann-Langevin approach to the
classical Boltzmann simulations performed by
Molnar \cite{Molnar:2004ph} and subsequent simulations of  
Zhang {\it et al.\ }\cite{Zhang:2005ni}. 
We will concentrate on Molnar's simulation here.
In this simulation  the gluon-gluon cross section 
takes the form, $\frac{d\sigma}{dt} \propto 1/(t-\muD^2)^2$, with
screening mass, $\muD=0.7\,\mbox{GeV}$. 
The total gluon-gluon cross section fixes $\alphas$ through
the relation $\sigma_0 = \frac{9 \pi\alphas^2}{2 \mu_D^2}$. The constants
are determined by comparing the model cross sections at finite $t$ 
and zero $\muD$  
with the corresponding unscreened perturbative 
expressions in QCD. The extracted $\alphas$ is 
used to set the charm-quark cross sections.
To estimate the diffusion coefficient in this model
we follow Appendix \ref{nonrelapp}, suitably modified
for the classical statistics in Molnar's simulation.
The diffusion coefficient of the charm quark in this model  can
be written in the follow form:
\begin{equation}
\label{Dmolnar}
    D = \frac{1}{(\cf/\ca)\, n_0 \sigma_0} \, 
      f\left(\frac{\muD}{T}, \frac{n_q}{n_g}\right) \; ,
\end{equation}
where $n_g$ is the density of gluons, $n_q$ is the 
density of quarks plus anti-quarks, $n_0=n_g + n_q$ is
the total particle density, and the 
function $f(\muD/T, n_q/n_g)$  is determined
after Appendix \ref{nonrelapp}. $f(\muD/T, n_q/n_g)$ is
approximately two for typical simulation parameters.
   
The magnitude of the resulting elliptic flow in Molnar's classical Boltzmann
simulation, with $\frac{dN}{dy}=2000$ for central collisions
and $\sigma_0=10\,\mbox{mb}$,
is similar to our results with  $D \approx 3/(2\pi T)$. 
Since elliptic flow develops over a period of 1--4\,fm 
we will estimate the diffusion coefficient at
$\tau \approx 2\,\mbox{fm}$ in Molnar's simulation and compare it
with $D\approx 3/(2\pi T)$. 
The precise value of Molnar's diffusion coefficient depends 
on $\muD/T$,  which we take to be $2.5$. 
 It also depends on the ratio of quarks to gluons 
which we take to be $n_q/n_g=36/16$ for chemically equilibrated 
3-flavor matter and $n_q/n_g=0$ for gluon dominated matter.
Using Glauber calculations, we estimate the transverse
area for central collisions to be $A = 100\,\mbox{fm}^2$.  
Then, $n_0 = (dN/dy)/(\tau A)$.
With these inputs, the diffusion coefficient in Molnar's model is 
\begin{equation}
\label{MolnarD}
    \left. D \right|_{\tau \approx 2\,{\rm fm}} 
     \approx 0.225 \, \mbox{fm} \left\{ 
\begin{array}{ll}
     1.4
    & \qquad \mbox{(gluon dominated)} \\ & \\ 
     2.5
                  & \qquad \mbox{(chemically equilibrated)}
\end{array} \right. \;  .
\end{equation}
Taking $T\approx 200\,\mbox{MeV}$ 
and $D \approx 3/(2\pi T)$, the diffusion coefficient 
in our model is 
\begin{equation}
\label{OurD}
     \left. D \right|_{\tau \approx 2\,\rm{fm}} \approx 0.47\, \mbox{fm}  \; .
\end{equation}
Comparing \Eq{MolnarD} and \Eq{OurD} we conclude that both approaches agree on the magnitude of 
the elliptic flow when the transport coefficients are comparable. Our
simulation with $D=3/(2\pi T)$ corresponds to Molnar's simulation
with $dN/dy=2000$ for central collisions and $\sigma_0=10\,\mbox{mb}$.

\section{Discussion}

Fig.~\ref{RAA}(a) and (b) summarize the phenomenological application
of our work. Several items are apparent in these figures.

First, we see in \Fig{RAA}(a) that there is no significant suppression in
$R_{AA}$ until the diffusion coefficient is of order,
$D \approx 20/(2\pi T)$.  This diffusion coefficient should
be compared with the perturbative estimate made in
\Sect{diffusion:sect},
\begin{eqnarray}
    D \approx \frac{6}{2\pi T} \left( \frac{0.5}{\alphas} \right)^2 \; .
\end{eqnarray}
The suppression seen in \Fig{RAA}(a) at 
large transverse momentum, $p_{T}\gsim 4.0\,\mbox{GeV}$, is 
somewhat overestimated since the Langevin description 
underestimates the
longitudinal fluctuations by a factor of three to four in this 
momentum range. However, for $p_{T} \gsim 4.0\,\mbox{GeV}$
radiation should become important  and further suppress the 
spectrum. 

Examining \Fig{RAA}(a) at large momentum, we see that 
the suppression decreases only slightly 
for $p_{T}$ greater than $\approx 3\,\mbox{GeV}$. 
This is reminiscent of the suppression pattern seen in
the light hadron data. It seems that all calculation which
have a reasonably complete model of the fluctuations can reproduce
this trend \cite{Jeon:2003gi,Molnar:2004zj}.
This suppression pattern is approximately independent of the model
for the transport coefficients. Looking at Fig.~\ref{RAA2}(a), we see
essentially the same pattern when $\frac{dp}{dt} \propto v$.  However,
in this case the magnitude of the suppression is significantly smaller
for a given value of the diffusion coefficient. 

Turning to elliptic flow, we see in Fig.~\ref{RAA}(b) that 
the
heavy quark elliptic flow rises with
momentum until $p_{T}\approx 1-2\, \mbox{GeV}$, and then flattens. 
(When $D=1.5/(2\pi T)$, $v_2(p_T)$ shows a qualitatively different behavior
which will be discussed shortly.)
A similar transition is seen in the elliptic flow of light hadrons,
illustrated by the band in the figure. For the light hadron data,
it has been argued that the
transition from rising to flat is controlled by transport coefficients,
and signals a transition from a quasi-thermal to a kinetic regime
\cite{Molnar:2001ux,Teaney:2003pb}. A similar
argument can be given here. Examining Fig.~\ref{RAA}(a),  we
see that $R_{AA}$ has a funny shape at low momentum
and then starts to fall exponentially. These features reflect
the thermal distribution.
At higher momentum, $R_{AA}$ stops falling exponentially and approaches
a constant. This transition coincides with the transition seen in
$v_2(p_T)$ and
depends on the diffusion coefficient. A detailed study of the
correlation between the initial angle of the charm quark and the
final angle of the charm quark confirms the qualitative conclusion
that the changes seen in $v_2$ and $R_{AA}$ for
$p_{T}\approx 1-2\,\mbox{GeV}$ reflect a transition
from a quasi-thermal to a kinetic regime. Only models which
smoothly interpolate between these two regimes can effortlessly reproduce
these trends in the entire $p_T$ range.

It has been argued that the charm quark could participate in the flow, and thus
be pushed out to higher transverse momentum \cite{Batsouli:2002qf,Greco:2003vf}. This scenario depends on the
diffusion coefficient. A detailed analysis of the initial and final momenta of
the charm quarks shows that, unless the diffusion coefficient is $\approx
1.5/(2\pi T)$ or less, the charm quarks are not pushed to higher momentum by
the flow. Indeed, when the diffusion coefficient is greater  than $1.5/(2\pi
T)$,\, a particle with final momentum $1.8$ GeV typically started with
momentum larger than this value. When the diffusion coefficient is small, $D\lsim
1.5/(2\pi T)$, the heavy quarks are pushed out to higher momentum by the flow.
This push 
is reflected in the $p_T$ dependence of the elliptic flow, as illustrated
in \Fig{RAA}(b). 
The rapid  fall of $v_2(p_T)$ for $p_T\approx 3.0\,\mbox{GeV}$ indicates that
the hydrodynamics is not  able to push the heavy quarks  beyond 
this transverse momentum for this diffusion coefficient.

The $p_{T}$ dependence of the elliptic flow also reflects aspects of the
underlying QCD matrix elements. In contrast to a constant energy loss model,
where $\frac{dp}{dt} \propto v$, the LO QCD  drag coefficient rises with the
momentum of the heavy quark (see \Fig{transportc}(a)).  Consequently, 
in the constant energy loss model, the elliptic flow decreases steadily at large
momentum rather than remain constant as in the LO QCD model. The elliptic 
flows of the two models are compared in \Fig{RAA}(b) and \Fig{RAA2}(b).

Examining Fig.~\ref{RAA}(a) and (b), we reach the important conclusion that
$R_{AA}$ and $v_2(p_T)$ are tightly correlated.  At low momentum, this
correlation is encoded in Einstein's relation between the energy loss and the
hydrodynamic diffusion coefficient, 
\[
\left.\frac{dE}{dx}\right|_{v\approx0} = \frac{T}{D} \; .  
\] 
It is impossible to have a large elliptic flow without
also predicting a significant suppression of charm quarks.  Indeed, if future
measurements find the  charm elliptic flow  strong, and the spectrum 
not suppressed, the present authors must logically conclude that
hydrodynamics is not responsible for the observed elliptic flow. This strong
conclusion might be mitigated if coalescence or some other exotic hadronization
mechanism manages
 to make the $D$ meson elliptic flow significantly larger than
the underlying $c$-quark elliptic flow. 

This point is underscored in recent classical Boltzmann simulations by
Molnar \cite{Molnar:2004ph} and subsequent simulations by 
Zhang {\it et al.\ }\cite{Zhang:2005ni}.  We found in 
\Sect{comparison} that our charm quark elliptic flow is
comparable to Molnar's results, provided  the two
simulations have approximately the same diffusion coefficient. Our
simulation with $D=3/(2\pi T)$ corresponds to his simulation with
$dN/dy=2000$ for central collisions and $\sigma_0=10\,\mbox{mb}$.
However, in Molnar's work, the
final D-meson elliptic flow is significantly larger than the underlying
c-quark flow  as a result of coalescence.  
Similarly in the simulation of Zhang {\it et al.}, 
the final D-meson elliptic flow is amplified by coalescence, although
the amplification factor is smaller than in Molnar's work due to
the details of the coalescence model. Whether coalescence can amplify the
elliptic flow of D-mesons in a complete dynamical setting remains 
unclear \cite{Molnar:2004zj,Molnar:2004rr}.

In summary, we have calculated various transport coefficients for a heavy
fermion in the perturbative Quark Gluon Plasma.   Since the gamma factor of the
charm quark  is not particularly large in much of the experimental momentum range,
$\gamma v \lsim 4$,  collisions rather
than radiation should determine the medium modifications of the heavy quark
spectrum.  Therefore, we have re-examined collisional energy loss, and also
calculated  momentum broadening, which is essential to determine how the heavy
quark spectrum will be  modified.  To quantitatively asses the modifications of
the heavy quark spectrum, we formulated a Boltzmann-Langevin model for
the heavy quark kinetics. The model is strictly valid  for
non-relativistic quarks and
for relativistic quarks to leading log of $T/m_D$.  We characterized the
strength of the energy loss and momentum diffusion in terms of the diffusion
coefficient, which in perturbation theory is approximately, $D \times (2 \pi T)
\approx 6\,(0.5/\alphas)^2$.  The Boltzmann-Langevin model was solved
analytically for a Bjorken expansion, giving a simple estimate
shown in \Fig{distribution} for the medium modifications of the heavy
quark spectrum.
Then we solved the Langevin equations numerically, yielding
the modification factor $R_{AA}$ and $v_2(p_T)$  for charm quarks.
The results are shown in \Fig{RAA}(a) and (b).

The diffusion coefficient for a relativistic plasma
is of order $\sim \tau_{c}$, where $\tau_{c}$ is the typical
collision time of the heavy quark. Thus, future measurements
at RHIC will produce a tantalizing  estimate of this
fundamental time scale. Since the diffusion coefficient is
a well defined concept, this estimate might ultimately be
compared to Lattice QCD calculations, confirming or rejecting the
paradigm of QGP formation in relativistic heavy ion collisions.

\noindent {\bf Acknowledgments.} We thank Peter Petreczky for
useful discussions. This work was supported by grants
from the U.S. Department of Energy, DE-FG02-88ER40388 and DE-FG03-97ER4014, 
from the Natural Sciences and Engineering Research Council of Canada, and from 
le Fonds Qu\'eb\'ecois de la Recherche sur la Nature et les Technologies.
\appendix
\section{Solution of The Fokker-Plank equation fork
a Bjorken expansion}

\label{appendixBj}

Our starting point is the Boltzmann-Fokker Planck Equation (BFPE)
for the probability $P(\x,\p,t)$
to find a particle  at $(\x,\p)$,
\begin{equation}
\frac{ \partial P }{\partial t} + \frac{p^{i}}{M} \frac{ \partial P }
{\partial x^{i} } =
   \left[ \frac{ \partial} {\partial p^i}  \eta_D  p^i   +
   \frac{ \partial^2 } {\partial p^2 }   M T \;\eta_D  \right] P(\x,\p,t) \; .
\label{bfpe2}
\end{equation}
It is important to remember that
$P$ depends on space and time through the flow velocity and
temperature dependences of the medium.

For a Bjorken expansion, the probability distribution is independent of
the transverse coordinates and invariant
under boosts in the $z$ direction. This allows the l.h.s.\ of the BFPE
to be simplified \cite{Baym:1984np}.  Without loss of generality, we can
restrict our attention to mid-rapidity, $z=0$.
Demanding that the probability distribution be invariant
under small longitudinal boosts leads to the requirement that
\begin{eqnarray}
   t\, \left. \frac{ \partial P(z,\pperp, p_z) }{\partial z} \right|_{z=0} &=&
      -M \, \left. \frac{\partial P(z, \pperp, p_z)} {\partial p_z}  \right|_{z=0} \; ,
\label{derivs}
\end{eqnarray}
where we have used the non-relativistic approximation $E=M$, and where
$p^\perp_a$ denotes the transverse vector, {\it i.e.\ }
$p^\perp_a = (p_x, p_y)$. With the standard substitutions
\cite{Baym:1984np},
\begin{eqnarray}
\ov{p}_z  &\equiv& p_z  \frac{ t}{t_o} \; , \\
P(t, p_z) &\equiv& \ov{P}(t, \ov{p}_z ) \; ,
\end{eqnarray}
and \Eq{derivs}, the  BFPE reads,
\begin{equation}
\label{bfpeBj}
\frac{ \partial \ov{P}}{\partial t }  =
3 \eta_D \ov{P} + \eta_D \pperp_a \frac{ \partial \ov{P} }{\partial \pperp_a}
 +  \eta_D \ov{p}_z \frac{ \partial \ov{P} } { \partial \ov{p}_z }
 + M T \eta_D \frac{ \partial^2 \ov{P}} { \partial \pperp_a \partial \pperp_a}
 + M T \eta_D \left(\frac{t}{t_0}\right)^2
 \frac{ \partial^2 \ov{P}} { \partial \ov{p}_z ^2 } \; .
\end{equation}
This may be turned into the diffusion equation by employing the method of
characteristics \cite{Reif}.  With the following substitutions,
\begin{eqnarray}
     \uperp_a(t) &=& \pperp_a  e^{\int^t_{t_0} dt \; \eta_D } \; , \\
      u_z(t) &=& \ov{p}_z  e^{\int^{t}_{t_0} dt \; \eta_D } \; , \\
      \ov{P}(t, \pperp_a, \ov{p}_z) &=& f(t, u^{\perp}, u_z) \;
      e^{3 \int^t_{t_0}dt\;\eta_D} \; ,
\end{eqnarray}
\Eq{bfpeBj} becomes
\begin{eqnarray}
 \frac { \partial f } { \partial t } &=&
M T   \; \eta_D \;
e^{2\int^{t}_{t_0} dt\; \eta_{\scriptscriptstyle D} }   \;
\frac{ \partial^2 f} { \partial \uperp_a \partial \uperp_a}
+
M T \eta_D \left(\frac{t}{t_0}\right)^2  \;
e^{2\int^{t}_{t_0} dt\, \eta_D }   \;
\frac{ \partial^2 f} { \partial u_z^2} \; .
\end{eqnarray}
Employing separation of variables and one more change of variables, we write
\begin{eqnarray}
   f(t,\uperp, u_z) = f_x (\theta^{\perp}, u_x) \;  f_y(\theta^{\perp}, u_y) \; f_z(\theta_z, u_z) \; ,
\end{eqnarray}
with
\begin{eqnarray}
  \theta^{\perp}  &\equiv& \int_{t_0}^t \; M T \eta_D   \; e^{2\int_{t_0}^t dt \, \eta_D } \; ,\\
  \theta_{z}  &\equiv& \int_{t_0}^t \; M T \eta_D \left(\frac{t}{t_0}\right)^2
  \; e^{2\int_{t_0}^t dt  \eta_D } \; ,
\end{eqnarray}
and find three uncoupled diffusion equations,
\begin{eqnarray}
   \frac{ \partial f_x} {\partial \theta^{\perp} } &=&
   \frac{ \partial^2 f_x} {\partial u_x^2 } \; , \\
   \frac{ \partial f_y} {\partial \theta^{\perp} } &=&
   \frac{ \partial^2 f_y} {\partial u_y^2 } \; ,\\
   \frac{ \partial f_z} {\partial \theta_{z} }  &=&
   \frac{ \partial^2 f_z} {\partial u_z^2 } \; .
\end{eqnarray}
The Green function for the diffusion equation is well known,
\begin{equation}
f_x(\theta^\perp, u_x) = \frac{1}{\sqrt{ 4 \pi \theta^\perp} } e^{-\frac{u_x^2}{4\theta^{\perp} } }  \; .
\end{equation}
Unraveling the nested definitions, we find the following Green function
for the BFPE for a Bjorken expansion:
\begin{eqnarray}
P_z(\p, t | \p_0,  t_0)
 &=&
 \frac{ 1}{\sqrt{ 2 \pi M T^{\perp}(t) }  }
 \exp\left[ - \frac{ (p^x - p_0^x e^{-\chi(t)})^2 }{2 M T^{\perp}(t) } \right] \nonumber \\
 &\times&
 \frac{ 1}{\sqrt{ 2 \pi M T^{\perp}(t) }  }
  \exp\left[ - \frac{ (p^y - p_0^y e^{-\chi(t)})^2 }{2 M T^{\perp}(t) } \right]
 \nonumber  \\
  &\times &
 \frac{t_0}{t} \frac{ 1}{\sqrt{ 2 \pi M T^{z}(t) }  }
 \exp\left[ - \frac{ (p^z - \left(\frac{t_0}{t}\right)\; p_0^z \; e^{-\chi(t)})^2 }{2 M T^{z}(t) } \right] \; ,
\end{eqnarray}
with the intuitive definitions:
\begin{eqnarray}
\label{teff2}
   \chi(t) &=&  \int_{t_0}^{t}dt' \;\eta_D(t') \; ,  \\
   T^\perp(t) &=&   2\int_{t_0}^{t}dt' \;\eta_D(t')  \; T(t') \;
  e^{2 \chi(t')  - 2\chi(t) } \; ,\\
   T^{z}(t) &=&   2\int_{t_0}^{t}dt' \;\eta_D(t') \;  T(t')
   \left( \frac{t'}{t} \right)^2
  e^{2 \chi(t')  - 2\chi(t) } \; .
\end{eqnarray}
Switching from $z$ to $\eta$ coordinates, we define
$P(\p,t|\p_0,t_0) \equiv (t/t_0) P_z(\p,t|\p_0,t_0)$ and
have
\begin{eqnarray}
\label{convolve-app}
 \left. \frac{dN}{d^3p\, d\eta} \right|_{\eta=0,t}
  =  \int d^3p_0 P(\p,t | \p_0,t_0)
           \, \left. \frac{dN}{d^3p_0\, d\eta} \right|_{\eta=0,t_0} \; .
\end{eqnarray}
The differences between the transverse and longitudinal directions
of the Green function should be noted.

Now let us substitute the form expected for the Bjorken solution into
these equations. Noting that $\eta_D\propto T^{2}$, we have
\begin{eqnarray}
          T(t) &=&  T(t_0) \left( \frac{t}{t_0} \right)^{-\alpha } \; ,\\
     \eta_D(t) &=&  \eta_D(t_0) \left( \frac{t}{t_0} \right)^{- 2\alpha }  \; .
\end{eqnarray}
Defining $\chi_0 \equiv \eta_D(t_0)\; t_0 /(1-2\alpha)$ and
changing integration
variables to integrate over $\chi$ produces
\begin{eqnarray}
   \chi &=&  \chi_0     \;
   \left[ \left(\frac{t}{t_0}\right)^{1-2\alpha}  - 1 \right]  \; , \\
   T^\perp(t) &=&  T(t)\; h_{\frac{-\alpha}{1-2\alpha} }(2\chi + 2\chi_0 )  -
      e^{-2\chi}  \;
      T(t_0) \;
      h_{\frac{-\alpha}{1-2\alpha} } (2\chi_0) \; ,\\
   T^z(t) &=&  T(t)\;
              h_{\frac{2-\alpha}{1-2\alpha} } (2\chi + 2\chi_0 )  -
               e^{-2\chi} \; T(t_0) \;
               \left(\frac{t_0}{t}\right)^2
               \; h_{\frac{2 - \alpha}{1-2\alpha}} (2\chi_0) \; .
\end{eqnarray}
Here we have defined the function
\[
h_{\gamma}(x) = \left\{ \begin{array}{ll}
x^{-\gamma} \; e^{-x} \;\int_0^{x} dx' \; x'^{\gamma} e^{x'} \qquad
                     &    \gamma>-1  \\ & \\
x \; e^{-x} \; \mbox{Ei}(x)  &     \gamma=-1
\end{array} \right. \; ,
\]
and employed the exponential integral $\mbox{Ei}(z) = -\int_{-z}^{\infty}dt\; e^{-t}/t$.\,
$h_{\gamma}(x)$ may be expressed in terms of the incomplete
gamma function, but we did not find the results useful in
developing the series expansions below.  For small $x$ we have
\begin{equation}
h_\gamma(x) \approx \left\{
\begin{array}{ll}
  \frac{x}{\gamma+1} - \frac{x^2}{(\gamma+1)(\gamma+2)} + ... &
\gamma>-1  \\ & \\
(\ln(x) + \gamma_{E})\;x   - (-\ln(x) -\gamma_E +1)\;x^2 + ... \qquad
  & \gamma=-1
\end{array} \right. \; ,
\end{equation}
where $\gamma_E \approx 0.5772$ is the Euler-Mascheroni constant.
For large $x$ and all values of $\gamma$  we have
\begin{eqnarray}
   h_{\gamma}(x)  &\approx& 1 - \frac{\gamma}{x} +
   \frac{\gamma(\gamma-1)}{x^2} + ...  + O(e^{-x}) \; .
\end{eqnarray}

For small values of $\chi$ and $\chi_0$
the heavy quark spectrum is only slightly affected by the
surrounding medium.
Using the small argument expansion of $h_\gamma(x)$,
we have for small values of $\chi$ and $\chi_0$\ ,
\begin{equation}
T^\perp (t) = T(t_0)\: 2\eta_D(t_0) t_0 \times \left\{
\begin{array}{ll}
   \frac{1}{1 - 3\;\alpha}
   \left[\left(\frac{t}{t_0}\right)^{1 - 3\alpha} -1\right]  \;
    & \qquad \alpha  < \frac{1}{3}  \\ & \\ 
   \ln \left( \frac{t}{t_0} \right) \;
                  & \qquad \alpha = \frac{1}{3}
\end{array} \right. \; .
\end{equation}
Similarly, for $T_z$ we have
\begin{eqnarray}
   T^{z}(t) = T(t_0) \; 2\eta_D(t_0) t_0 \; \left(\frac{t_0}{t}\right)^2\; \frac{1}{3 - 3\alpha}
   \left[ \left(\frac{t}{t_0}\right)^{3-3\alpha} - 1 \right] \; .
\end{eqnarray}
For small $\chi$ the
effect of rescattering is to convolute the original transverse momentum distribution with a Gaussian of transverse width $M T^\perp(t)$  and
the original longitudinal momentum distribution with a Gaussian of width
$M T^{z}(t)$.

For large values of $\chi$ the system approaches equilibrium and memory
of the original distribution is lost. The final distribution is a Gaussian
which is quite close to the thermal distribution. In this limit the effective temperatures
in the transverse and longitudinal directions are
\begin{eqnarray}
    T^{\perp}(t_f) &=& T(t_f) \left( 1   + \frac{\alpha }{2\eta_D(t_f)
    t_f} \right) \; ,  \\
    T^{z}(t_f) &=& T(t_f) \left( 1   - \frac{2-\alpha}{2\eta_D(t_f) t_f}
    \right)  \; . 
\end{eqnarray}
Thus, the mean squared transverse momentum  is slightly  {\it larger} than
the equilibrium  expectation  by an amount of order $1/(2\eta_D(t_f) t_f)$.
The longitudinal momentum is slightly {\it smaller} than the
equilibrium value by a similar amount.
These corrections are reminiscent
of viscous corrections to the stress energy tensor.

\section{Details of leading-order calculations}

\label{appendixA}

In this appendix we provide the complete details of the derivations of
the momentum diffusion and energy loss coefficients discussed in the
main text.

\subsection{Matrix elements}

We begin by proving that the matrix elements given in \Eq{eq:Msq1} are
correct.
The vacuum matrix elements are most easily found in the rest frame of
the heavy quark where  only $t$-channel gluon exchange contributes.
The external gluon polarization sum is performed with two
transverse polarization vectors which are purely spatial in this frame.
The amplitude for gluon exchange is $\sim P\cdot K/Q^2$
while the amplitude for a Compton process involves
\st
\M_{\rm Compton} \simeq \bar u(p) \nott{\epsilon} (-i\nott P -i\nott K -
M)^{-1} \nott{\epsilon}' u(p)
\sim \Tr (-i\nott{P}+m) \nott{\epsilon} \frac{-i\nott P -i\nott K +M}
{(P+K)^2 + m^2} \nott{\epsilon}' \; .
\stp
Since $P^\mu$ is purely temporal,\ $\epsilon$ anti-commutes across it,\
leading to $\M_{\rm Compton} \sim (P \cdot K)/(P\cdot K)$, which is
smaller by a factor of $k/M$ than the gluon exchange amplitude.
Therefore, the Compton amplitude can be ignored.

Up to a color factor, the squared matrix elements for the scattering of a heavy fermion
with a quark or gluon via t-channel exchange
can be written in the following way:
\st
\label{Me2GG}
\left|\M \right|^2 \propto L_{\mu \nu}(P) \,M_{\alpha \beta}(K,K') \, G^{\mu \alpha}(Q) G_{\nu \beta}(Q) \; .
\stp
Here $G^{\mu \alpha}(Q)$ is the internal gluon propagator which we will
take to be in Feynman gauge $\eta^{\mu \alpha}/Q^2$.  Tracing
over the heavy fermion line, one easily finds
\st
L_{\mu \nu}(P) = 4 P_\mu P_\nu \; .
\stp
This result is independent of the spin of the heavy particle
and depends only on the approximation that its mass is much larger than
the $Q^2$ of the process.  For scattering with a light quark, tracing
over the light fermion line yields
\st
4P_\mu P_\nu M^{\mu \nu} =
	 16 M^2 k^2 (1+\cos \theta_{kk'}) \; ,
\stp
where $\k$, $\k'=\k-\q$ are the bath particle momenta before and after
the collision.  For scattering with a gauge boson,
contracting the three gluon vertex with the polarization vectors yields
\st
 4 P_\mu P_\nu M^{\mu \nu} = 4 \left[ P^\mu ( (K_\mu+K'_\mu) \epsilon \cdot \epsilon'
	- \epsilon_\mu (K-Q)\cdot \epsilon'
	- \epsilon_\mu ' (K'+Q) \cdot \epsilon ) \right]^2 \; .
\stp
Using $\epsilon \cdot P=\epsilon' \cdot P =0$, we have
\st
 4 P_\mu P_\nu M^{\mu \nu} = 16 M^2 k^2 (\epsilon \cdot \epsilon')^2
= 16 M^2 k^2 (1 + \cos^2 \theta_{kk'}) \; .
\stp

Next we find a covariant expression for the rest-frame angle $\cos \theta_{kk'}$.  Since $Q$ is
purely spatial and $K$ and $K'$ have the same energy in this frame, we
have that $\k'=\k-\q$ and $K'{}^2=0=K^2-2K\cdot Q+Q^2$.
Further, $K\cdot K'=k^2(1-\cos\theta_{kk'}) = K\cdot(K-Q)=-K\cdot
Q=-Q^2/2$.  Finally, $k^2 = (K\cdot P)^2/M^2$.
Therefore,
\st
\cos\theta_{kk'} = 1 - \frac{Q^2 M^2}{2(K\cdot P)^2} \; .
\stp
Using this, along with $M^2 k^2 = (K\cdot P)^2$, allows us to
covariantize the expressions we just found, resulting in
\Eq{eq:Msq2}.

\subsection{Momentum Diffusion -- Non-Relativistic Case}
\label{nonrelapp}

Next complete the calculation of \Sect{diffusion:sect}.
The matrix elements are screened replacing the $t$-channel
gluon propagator in \Eq{Me2GG} with the HTL
propagator.
This is most easily done in Coulomb gauge, where the
HTL propagator in the plasma frame is
\st
\label{HTL}
G_{\mu \nu}(Q) = \frac{- \delta_{\mu 0}\delta_{\nu 0}}{q^2 + \Pi_{00}}
	+ \frac{\delta_{ij} - \hat q_i \hat q_j}{q^2 - \omega^2
	+ \PiT} \; .
\stp
The substitution $\mu \rightarrow i$ indicates that only the
spatial parts participate.
Explicit expressions for
the HTL self-energies are \cite{Kalashnikov:1979cy,Weldon:1982aq}
\begin {eqnarray}
    \PiT(\omega,\q) & = & \mD^2
	\left\{
	    \frac{\omega^2}{2q^2}
	+
	    \frac{\omega \, ( q^2 {-} \omega^2 )}{4 q^3}
	    \left[
		\ln \left( \frac{q+\omega}{q-\omega}\right) - i \, \pi
	    \right]
	\right\}
	\; ,
\label{eq:PiT} \\
    \Pi_{00}(\omega,\q) & = & \mD^2
	\left\{
	    1
	    -
	    \frac{\omega}{2q}
	    \left[
		\ln \left(\frac{q+\omega}{q-\omega}\right) - i \, \pi
	    \right]
	\right\} \; .
\label{eq:PiL}
\end {eqnarray}

Since heavy quark is approximately at rest in the rest frame of the plasma
$P^{\mu}=(M, 0)$, only the temporal part  of the gauge boson
propagator $G^{00}(Q)$ enters when computing the squared matrix
elements in \Eq{Me2GG}.
For the case of momentum diffusion studied here the energy transfer $\omega$ is
small  and the HTL correction reduces to simple Debye screening,
$1/Q^2 = 1/q^2 \to 1/(q^2 + \mD^2)$.  With these observations and
the results of the preceding section,
the screened matrix elements \Eq{eq:Msq1} are reproduced.

The total momentum diffusion constant is then,
\bg
\label{eq:kappa-app}
3\kappa & = & \frac{1}{2M} \int \frac{d^3 \k d^3 \k' d^3 \q}{(2\pi)^9
	8 k^0 k'{}^0 M} (2\pi)^3 \delta^3(\k' - \q -\k) 2\pi
	\delta(k'-k) \q^2 \times \nonumber \\
	&& \qquad \qquad \times
	\left[ N_f |\M|^2_{\rm quark} n_f(k) (1{-}n_f(k'))
	+|\M|^2_{\rm gluon} n_b(k) (1{+}n_b(k')) \right] \; .
\nd
We use the 3-momentum delta function to perform the $\k'$
integration and rewrite the
the $\q$ integration  as
\begin{equation}
\int \frac{d^3 \q}{(2\pi)^3} = \frac{1}{4\pi^2} \int q^2 dq
\int_{-1}^1 d \cos \theta_{\k\q} \; .
\end{equation}
Using
the energy delta function to perform the $\cos \theta_{\k\q}$
integral  and noting the relation, $\cos \theta_{\k\k'}=1 - q^2/(2k^2)$, we have
\begin{eqnarray}
\label{eq:raw_kappa}
3\kappa & = & \frac{\ch g^4}{4\pi^3} \int_0^\infty k^2 dk \int_0^{2k} q dq
	\; \frac{q^2}{(q^2 + \mD^2)^2} \times \nonumber \\
	&& \qquad \times \left[
	N_f \frac{e^{k/T}}{(e^{k/T}+1)^2}
	\left( 2 - \frac{q^2}{2k^2} \right)
	+ N_c \frac{e^{k/T}}{(e^{k/T}-1)^2}
	\left( 2 - \frac{q^2}{k^2} + \frac{q^4}{4k^4} \right) \right]
	\; .
\end{eqnarray}
In performing the $q$ integral, we may drop resulting terms which are
subleading in $\mD/k$, because the dominant contribution is from $k
\gsim T$, where this ratio is small.  The $q$ integral in the small
$\mD/k$ limit gives
\begin{equation}
3\kappa = \frac{\ch g^4}{4\pi^3} \int_0^\infty k^2 dk
	\left[ \log \frac{4k^2}{\mD^2}-2 \right] \left[
	N_f \frac{e^{k/T}}{(e^{k/T}+1)^2}
	+ N_c \frac{e^{k/T}}{(e^{k/T}-1)^2}
	\right]
	\; .
\end{equation}
The remaining $k$ integral is straightforward and yields
\begin{equation}
3\kappa = \frac{C_H g^4 T^3}{6\pi} \left[
	N_c \left( \log \frac{2T}{\mD} + \frac{1}{2} - \gamma_E
	+ \frac{\zeta'(2)}{\zeta(2)} \right)
	+\frac{N_f}{2} \left( \log \frac{4T}{\mD} + \frac{1}{2} - \gamma_E
	+ \frac{\zeta'(2)}{\zeta(2)} \right) \right] \; .
\label{eq:kappa}
\end{equation}
Finally, the relation between $\kappa$ and the diffusion
coefficient can be used to determine the diffusion coefficient
given in the text.

Finally, let us indicate how \Eq{eq:raw_kappa} may be used 
to calculate the diffusion coefficient of a heavy quark
in Molnar's parton cascade model. In this model, $\alphas$ is related 
to the total gluon-gluon cross section via,  
$\sigma_0=\frac{4\pi C_A^2 \alpha_s^2}{d_A \muD^2}$,  where $d_A=N_c^2-1$ is 
the dimension of the adjoint representation, and $\muD$ is a
fixed Debye mass. The particles obey 
classical statistics and therefore $f(1\pm f)$ is replaced
by $Ce^{-p/T}$, where the constant $C$ is adjusted to 
reproduce the density of the classical particles for each species. With these
modifications \Eq{eq:raw_kappa} becomes
\begin{eqnarray}
\label{eq:raw_kappam}
3\kappa & = & \frac{\sigma_0\,\muD^2}{2\, T^3} \int_0^\infty k^2 dk \int_0^{2k} q dq
	\; \frac{q^2}{(q^2 + \muD^2)^2} \times \nonumber \\
	&& \qquad \times \left[
   \left(\frac{C_F}{C_A}\right)^2 n_q e^{-k/T}
	\left( 2 - \frac{q^2}{2k^2} \right)
	+ \frac{C_F}{C_A}\, n_g e^{-k/T}
	\left( 2 - \frac{q^2}{k^2} + \frac{q^4}{4k^4} \right) \right]
	\; .
\end{eqnarray}
where $n_q$ is the density of quarks plus anti-quarks and $n_g$ is
the density of gluons. The total particle density is $n_0=n_q + n_g$. 
The form for the diffusion coefficient given in the text \Eq{Dmolnar} follows from
\Eq{eq:raw_kappam} and the relation between $\kappa$ and $D$.

\subsection{Phase Space}

In order to calculate the transport coefficients we
need to integrate over the phase space in \Eq{eq:fs}.
In this sub appendix we will adapt
the integration technology of Baym {\it et al.\ } \cite{Baymetal}
to the peculiarities of the heavy quark phase space.

The desired phase space integration domain is
\st
\frac{1}{2p^0} \int \frac{d^3 \k d^3 \k' d^3 \p'}{(2\pi)^9
	8k^0 k'{}^0 p'{}^0} (2\pi)^4 \delta^4(P+K-P'-K') \; .
\stp
We use the spatial delta functions to perform the $\p'$ integration and
shift variables to integrate over the momentum transfer $\q \equiv \k-\k'$ 
rather than $\k'$.  Define the angles $\cos \theta_{kq}$ and $\cos
\theta_{pq}$ to be the angles between the $\k$ and $\q$ vectors and the
$\p$ and $\q$ vectors, and $\phi_{q;pk}$ to be the angle between the
plane containing $q,p$ and the plane containing $q,k$.  Also note that
$p'{}^0 = \sqrt{M^2 + (\p+\q)^2}$.  Taking $\p$ large, and noting that
$v \equiv p/p^0$ is the velocity, this becomes $p'{}^0=p^0 + qv\cos
\theta_{pq}$ plus terms of order $q^2/p^0$ which are small.  In the
denominator, the approximation $p'{}^0=p^0$ is good enough.

The phase space becomes,
\st
\frac{1}{16 (p^0)^2} \frac{1}{(2\pi)^3} \int_0^\infty \frac{kdk q^2 dq}{k'}
	\int_{-1}^1d\cos \theta_{pq}\; d\cos\theta_{kq}
	\int_0^{2\pi} \frac{d\phi_{q;kp}}{2\pi}
	\delta(p^0-p'{}^0+k-k') \; .
\stp
Now we introduce the integration variable $\omega$
\st
1 = \int d\omega \,\delta(\omega - p'{}^0 + p{}^0) \; ,
\stp
which is the energy transferred to the heavy particle.  We now
perform the two $\cos \theta$ integrations
using kinematic relations
to rewrite  the delta functions  as
\bg
\label{eq:kinematics}
\delta(\omega+p-p') & = & \frac{1}{vq} \delta\left(\cos\theta_{pq}
	-\frac{\omega}{qv} \right) \; ,
	\nonumber \\
\delta(\omega+k'-k) & = & \frac{k'}{kq} \delta\left( \cos\theta_{kq}
	- \frac{\omega}{q} + \frac{\omega^2-q^2}{2kq}\right)
	\Theta(k-\omega) \; .
\nd
The phase space integration becomes
\st
\frac{1}{16(p^0)^2} \frac{1}{(2\pi)^3} \int_0^\infty dq
	\int_{-vq}^{vq} \frac{d\omega}{v}
	\int_{\frac{\omega+q}{2}}^\infty dk
	\int_0^{2\pi} \frac{d\phi_{q;kp}}{2\pi} \; .
\label{eq:phase_space}
\stp
When $q$ is small, the lower limit on $k$ can be taken to 0 and
$\cos\theta_{kq}\simeq \omega/q$.

\subsection{Energy Loss and Momentum Diffusion -- Relativistic Case}

Next we will complete the calculation of energy loss and momentum diffusion
of \Sect{eloss-sect}.
The procedure is straightforward though somewhat cumbersome.
We calculate the matrix elements with
the HTL propagator, substitute the matrix elements into the
expressions for the transport coefficients (\Eq{eq:fs}), and finally integrate 
over the phase space numerically, as parametrized above.

First we write the matrix elements by inserting the HTL
propagator \Eq{HTL} into \Eq{Me2GG} and rewrite the
resulting expression in terms of the integration variables
introduced in the previous sub appendix.
Generally, the matrix elements
squared are of the form: $\left| \M \right|^2 = A + B\,\cos(\phi_{q:kp}) +
C\,\cos^2(\phi_{q;kp})$. After averaging over this azimuthal angle the  matrix
element becomes, $\llangle \M^2 \rrangle_{\phi} = A + C/2$.
For a light quark scattering off a heavy quark,
$\llangle \M^2 \rrangle_{\phi}$ becomes
\st
   \left[\mbox{cf}\right]\,16 (p^0)^2 \,
   \left(
	\frac{4k(k-\omega) - [q^2-\omega^2]}{2\left|q^2 + \PiL\right|^2}
	 \,+ \,
	\frac{[4k(k-\omega) + (q^2+\omega^2)][v^2 q^2 - \omega^2][q^2-\omega^2]}
	{4q^4\left|q^2-\omega^2 +\PiT\right|^2} \right) \; ,
\label{eq:msq_fermion}
\stp
where $[\mbox{cf}]$ denotes the color factor,
$\left[2 C_{H} g^4/2 \right]$ as in \Eq{eq:Msq2}.

For gluon scattering,  essentially the
same procedure is followed. However, the result suffers
from ambiguities unless the coupling is really
small, as discussed in \cite{AMY6}.
  We will follow the prescription of that work
and write the gluon matrix element as the scattering of
a fictitious quark (with the color charge of a gluon)
plus an infrared finite remainder.
To this ephemeral quark
we will then apply the HTL corrections
and ignore  corrections to the finite piece.
This procedure
is correct to leading order since HTL corrections are only needed
for the eikonal vertex, which is independent of the spin
of the hard particle. This prescription does something reasonable
when $m_{D}/T$ becomes of order one.
Implementing this discussion, we compare the quark and gluon matrix elements in \Eq{eq:Msq2} and add
\begin{equation}
     [\mbox{cf}] \,16 \,
     \left( \frac{(p^0)^2 (v^2-1)}{2(q^2 -w^2)} + \llangle \frac{M^4}{4 (P\cdot K)^2} \rrangle_{\phi}  \right) \; ,
\end{equation}
to the quark expression, \Eq{eq:msq_fermion},
to obtain the gluon squared matrix element, $\llangle M^2 \rrangle_{\phi}$ .
The color factor $[\mbox{cf}]$ is $\left[ N_c C_{H} g^4 \right]$
for gluon case. We have not written out the $M^4/4 (P\cdot K)^2$ term
because it is awkward in this choice of integration variables
and will be integrated over separately.

To integrate the $M^4/4 (P\cdot K)^2$ term we
use a different parametrization of the phase space integrals.
We do not 
introduce $q$, but align $\p$ with the $z$ axis:
\bg
&&\int \frac{d^3 k d^3 k' d^3 p'}{(2\pi)^9 16 (p^0)^2 k^0 k'{}^0}
	(2\pi)^4 \delta^4(p+k-p'-k')
\\
&=& \frac{1}{16 (p^0)^2} \frac{1}{(2\pi)^3}
	\int_0 k\, dk\, k'\, dk' \int_{-1}^1 d\cos \theta_{kp}
	d\cos\theta_{k' p} \int_0^{2\pi} \frac{d\phi_{p;kk'}}{2\pi}
	\delta(k-k' +p^0-p'{}^0) \; . \nonumber
\nd
In these variables
$p'{}^0 = p^0 + v k \cos \theta_{kp} - v k' \cos \theta_{k'p}$.
Next we introduce a new integration variable
\st
1=\int_0^\infty d\omega\, \delta(\omega-k(1 - v \cos \theta_{kp})) \; ,
\stp
and use the two delta functions to perform the two angular
integrations, which yields 
\st
\frac{1}{16(p^0)^2} \frac{1}{(2\pi)^3} \int_0^\infty d\omega
	\int_{\frac{\omega}{1{+}v}}^{\frac{\omega}{1{-}v}}
	\frac{dk}{v} \frac{dk'}{v} \int_0^{2\pi} \frac{d\phi_{p;kk'}}{2\pi} \; .
\stp
The energy transfer is $q^0 = k'-k$ and
$P\cdot K = p^0 k(-1 + v \cos \theta_{pk}) = -\omega p^0$.  The vector
$\q$ is complicated in this coordinate system, but this vector is not needed to
integrate $M^4 / 4(P\cdot K)^2$.
With this parametrization, the $\phi_{p;kk'}$ integral over $M^4/4(P\cdot K)^2$  can immediately be performed, leaving three integrals that
are performed numerically.

In summary, we take the screened matrix elements
insert them into the expressions for the transport coefficients
and finally perform the phase space integrals. The results of
this procedure is illustrated in Fig.~\ref{transportc}  and has been discussed
in the main text.

Analytic expressions for the
transport coefficients can be derived 
to leading logarithm in $T/\mD$ \cite{Braaten:1991we}.  We will discuss the
energy loss coefficient
and leave the details of the other transport coefficients to
the reader. For $q \ll T$ the matrix elements
for the scattering of a light quark or gluon on a heavy quark are identical
up to a color factor
\begin{equation}
\llangle \M^2 \rrangle_{\phi} = [\mbox{cf}]\  16 \left( 2\frac{(p^0 k)^2}{|q^2 + \Pi_{00}|^2}
	+ \frac{(p^0 k)^2 (q^2 - \omega^2) (q^2 v^2 - \omega^2)}
	{|q^2 - \omega^2 + \PiT|^2} \right) \; .
\end{equation}
Again the color factor $[\mbox{cf}]$ is $[2 C_H g^4/2]$ for quarks and
$[\nc C_H g^4]$ for gluons.
For small $q$ and $\omega$ the weight factor in \Eq{eq:fs} is 
\begin{equation}
\label{weightf}
\frac{\omega}{2v} \left( n[k{-}\omega](1\pm n[k]) - n[k] (1\pm n[k{-}\omega]\right)
	\simeq \frac{\omega^2}{2vT} n[k](1 \pm n[k]) \; .
\end{equation}
Next we insert this weight factor and the small $q$  matrix elements
into \Eq{eq:fs} and \Eq{eq:phase_space} for the energy loss rate. After performing the
integral over $k$, the momentum lass rate is
\st
\label{lldpdt}
\frac{dp}{dt} = v \left( N_c + \frac{N_f}{2} \right)
	\frac{C_H g^4 T^2}{12\pi} \int dq \int_{-vq}^{vq}
	\frac{d\omega}{v} \frac{\omega^2}{2v^2}
	\left( \frac{q^4}{|q^2 + \Pi_{00}|^2}
	+ \frac{(q^2-\omega^2)(v^2 q^2-\omega^2)}
	{2|q^2 - \omega^2 + \PiT|^2} \right) \; .
\stp

To determine the leading-log coefficient, one should find the
coefficient of $dq/q$ when one drops the self-energies in this
expression. Following this prescription and performing the $\omega$ integral,
we transform \Eq{lldpdt} into
\bg
\frac{dp}{dt}
& \simeq &  v \left( N_c + \frac{N_f}{2} \right)
	\frac{C_H g^4 T^2}{24\pi} 
	\left( \frac{1}{v^2} - \frac{1{-}v^2}{2v^3}
	\log\frac{1{+}v}{1{-}v} \right) \, \log(T/\mD) \; .
\nd
The coefficient in front of $\log(T/m_D)$ is referred to as the leading-log
coefficient below.  A similar analysis but with different weight functions gives the
leading log expressions for the transverse and longitudinal momentum  diffusion
coefficients  given in the text, \Eq{eq:leadinglog}.

A procedure which is correct to next-to-leading logarithm was formulated by
Braaten and Yuan \cite{Braaten:1991dd} and then used for heavy quark energy loss by Braaten
and Thoma \cite{Braaten:1991we}.  Here the momentum integration is divided up into a soft
region $q < q^{*} $ and a hard region $q^{*} < q$, where  $q^{*}$ is some
intermediate scale $\mD \ll q^{*} \ll T$.
To determine the soft contribution to transport coefficient, one
should integrate \Eq{lldpdt} up to $q^{*}$ with the self
energies.
For $q^{*} \gg \mD$, this integral is  a number
$A_{\scriptscriptstyle \rm soft}(v)$  plus
the leading-log coefficient times $\log(q^*/\mD)$.
To determine the hard contribution to the
transport coefficient, one should drop the self-energies in \Eq{eq:msq_fermion} and
numerically integrate the matrix elements over the phase space with $q >
q^{*}$.  For $T \gg q^{*}$, this integral is a number 
$A_{\scriptscriptstyle \rm hard}(v)$,
plus  the leading-log coefficient times
$\log(T/q*)$. The sum of the hard and soft contributions is independent of
$q^{*}$ and determines the transport coefficient. The results of this
procedure is the following parametrization of the transport properties of a
heavy quark in the QGP
\begin{eqnarray}
\frac{dp}{dt} & =&
	\frac{C_H g^4 T^2}{24 \pi}\,v\,
	\left( \frac{1}{v^2} - \frac{1{-}v^2}{2v^3}
	\log\frac{1{+}v}{1{-}v} \right) \,
   \nonumber \\
    &  & \times \;\;
    \left[ N_c\left(\log(T/\mD) + A_b(v)\right)
    + \frac{N_f}{2}\left(\log(T/\mD) + A_f(v) \right) \right] \; ,
\label{eq:hqnll1} \\
  \kappa_T &=&
	\frac{C_H g^4 T^3}{12\pi}
	\left( \frac{3}{2} - \frac{1}{2v^2} + \frac{(1{-}v^2)^2}{4v^3}
	\log\frac{1{+}v}{1{-}v} \right) \,
   \nonumber \\
    &  & \times \;\;
    \left[ N_c\left(\log(T/\mD) + B_b(v)\right)
    + \frac{N_f}{2}\left(\log(T/\mD) + B_f(v) \right) \right] \; , \\
\label{eq:hqnll2}
\kappa_L & =&
	\frac{C_H g^4 T^3}{12 \pi}
	\left( \frac{1}{v^2} - \frac{1{-}v^2}{2v^3}
	\log\frac{1{+}v}{1{-}v} \right) \,
   \nonumber \\
    &  & \times \;\;
    \left[ N_c\left(\log(T/\mD) + C_b(v)\right)
    + \frac{N_f}{2}\left(\log(T/\mD) + C_f(v) \right) \right] \; .
\label{eq:hqnll3}
\end{eqnarray}
The coefficients $A(v)$, $B(v)$ and $C(v)$ for bosons and fermions
are tabulated as a function of velocity in Table ~\ref{table_ABC}.
\begin{table}
\begin{tabular}{|c|cccccc|} \hline
$\qquad v \qquad$ &
$\quad A_b(v) \quad$ & $\quad A_f(v) \quad$ &
$\quad B_b(v) \quad$ & $\quad B_f(v) \quad$ &
$\quad C_b(v) \quad$ & $\quad C_f(v) \quad$  \\ \hline \hline
0.05 & 0.0488 & 0.7420 & 0.0503 & 0.7440 & 0.0502 & 0.7437 \\
0.10 & 0.0567 & 0.7499 & 0.0617 & 0.7570 & 0.0620 & 0.7568 \\
0.15 & 0.0692 & 0.7624 & 0.0793 & 0.7774 & 0.0813 & 0.7782 \\
0.20 & 0.0861 & 0.7793 & 0.1026 & 0.8047 & 0.1081 & 0.8080 \\
0.25 & 0.1074 & 0.8006 & 0.1312 & 0.8383 & 0.1425 & 0.8464 \\
0.30 & 0.1331 & 0.8262 & 0.1648 & 0.8782 & 0.1851 & 0.8941 \\
0.35 & 0.1633 & 0.8564 & 0.2032 & 0.9240 & 0.2366 & 0.9519 \\
0.40 & 0.1981 & 0.8913 & 0.2462 & 0.9758 & 0.2981 & 1.0211 \\
0.45 & 0.2380 & 0.9312 & 0.2937 & 1.0334 & 0.3712 & 1.1036 \\
0.50 & 0.2834 & 0.9765 & 0.3459 & 1.0973 & 0.4580 & 1.2018 \\
0.55 & 0.3348 & 1.0280 & 0.4029 & 1.1677 & 0.5617 & 1.3195 \\
0.60 & 0.3933 & 1.0864 & 0.4653 & 1.2456 & 0.6871 & 1.4622 \\
0.65 & 0.4600 & 1.1532 & 0.5339 & 1.3321 & 0.8415 & 1.6383 \\
0.70 & 0.5371 & 1.2303 & 0.6101 & 1.4293 & 1.0369 & 1.8616 \\
0.75 & 0.6276 & 1.3208 & 0.6963 & 1.5408 & 1.2943 & 2.1566 \\
0.80 & 0.7368 & 1.4299 & 0.7969 & 1.6728 & 1.6549 & 2.5703 \\
0.85 & 0.8747 & 1.5678 & 0.9200 & 1.8374 & 2.2119 & 3.2097 \\
0.90 & 1.0641 & 1.7572 & 1.0848 & 2.0625 & 3.2366 & 4.3846 \\
0.95 & 1.3796 & 2.0728 & 1.3528 & 2.4390 & 6.0376 & 7.5840 \\  \hline
\end{tabular}
\caption{ The six functions of velocity  $A(v)$, $B(v)$, and $C(v)$ for
bosons ($b$) and fermions ($f$) are defined by Eqs.~(\ref{eq:hqnll1}--\ref{eq:hqnll3}) and parametrize the transport properties of a heavy particle in the QGP. }
\label{table_ABC}
\end{table}

\end{document}